\documentclass[12pt]{article}
\usepackage[utf8]{inputenc}
\usepackage[top=50pt,bottom=50pt,left=68pt,right=66pt]{geometry}
\usepackage{amsmath}
\usepackage{bm}
\usepackage{booktabs}
\usepackage{amssymb}
\usepackage[title]{appendix}
\usepackage{mathrsfs}
\usepackage{cite}
\usepackage{float}
\usepackage{xcolor}
\usepackage{multirow}
\usepackage{graphicx,caption,subcaption}
\usepackage[space]{grffile}
\usepackage{ upgreek }
\newcommand{\overbar}[1]{\mkern1.5mu\overline{\mkern-1.5mu#1\mkern-1.5mu}\mkern 1.5mu}
\renewcommand{\arraystretch}{1.3}
\definecolor{darkblue}{rgb}{0.1,0.1,.7}
\usepackage[breaklinks=true,linktocpage=true,
colorlinks=true,urlcolor=darkblue,linkcolor=darkblue,
citecolor=darkblue,pdfpagelabels=true,hypertexnames=true,
plainpages=false,naturalnames=false,]{hyperref}

\usepackage{datetime}
\newdateformat{monthyeardate}{%
  \monthname[\THEMONTH], \THEYEAR}
\date{\monthyeardate\today}

\def\cl{{\cal L}}

\def\co{{\cal O}}

\newcommand{\RM}[1]{\mathrm{#1}}

\begin{document}

\renewcommand{\arraystretch}{1.3}
\thispagestyle{empty}

{\hbox to\hsize{\vbox{\noindent\monthyeardate\today\\
Version to appear in ApJS}}}

\noindent
\vskip2.0cm
\begin{center}

{\Large\bf Meso-inflationary Peccei--Quinn symmetry breaking with non-minimal coupling}

\vglue.3in

Yermek Aldabergenov,${}^{a,}$\footnote{ayermek@fudan.edu.cn (corresponding author)} Ding Ding,${}^{a,}$\footnote{ding\_ding@fudan.edu.cn} Wei Lin,${}^{a,}$\footnote{frogelw@gmail.com} Yidun Wan${}^{a,b,}$\footnote{ydwan@fudan.edu.cn (corresponding author)}
\vglue.1in

${}^a$~{\it State Key Laboratory of Surface Physics, Center for Astronomy and Astrophysics, Department of Physics, Center for Field Theory and Particle Physics, and Institute for Nanoelectronic devices and Quantum computing, Fudan University,
 2005 Songhu Road, Shanghai 200433, China}\\
${}^b$~{\it Shanghai Research Center for Quantum Sciences, 99 Xiupu Road, Shanghai 201315, China}\\
\vglue.1in

\end{center}

\vglue.3in

\begin{center}
{\Large\bf Abstract}
\vglue.2in
\end{center}

We study a realization of the inflationary scenario where the Peccei--Quinn (PQ) symmetry is spontaneously broken during inflation, facilitated by its non-minimal coupling to gravity. This results in effectively two-field inflation: the early stage is driven by an inflaton field with the PQ symmetry intact, and the later stage is driven by the PQ scalar after its effective mass becomes tachyonic, causing destabilization from the origin. The non-minimal coupling serves the dual purpose of restoring the PQ symmetry during early inflation and flattening the PQ potential post-tachyonic shift, allowing for continued slow roll. We analyze the inflationary background solutions and scalar perturbations, which are amplified at small scales via significant isocurvature perturbations generated near the symmetry-breaking epoch. These perturbations lead to second-order gravitational waves, detectable by next-generation space-based experiments.

\newpage

\tableofcontents

\setcounter{footnote}{0}

\section{Introduction}

A natural solution to the strong CP problem of QCD is provided by the existence of a pseudo-Nambu-Goldstone boson, or axion, which couples to the gluon topological density, originating from the spontaneous breaking of a global abelian symmetry known as the Peccei--Quinn (PQ) symmetry. Additionally, the axion is one of the leading candidates for dark matter \cite{Preskill:1982cy,Abbott:1982af,Dine:1982ah,Sakharov:1994id,Sakharov:1996xg,Khlopov:1999tm}, motivating cosmological and astrophysical studies in this direction.

Along this line, we realize a meso-inflationary Peccei--Quinn Symmetry Breaking (PQSB) during the middle of observable inflation, facilitated by the non-minimal coupling of the Peccei--Quinn field to gravity. This has two key consequences:
\begin{enumerate}

    \item A two-stage inflation: the first stage driven by a PQ-neutral inflaton field with restored PQ symmetry, and the second stage driven by the PQ scalar after it becomes tachyonic, allowing for continued slow-roll inflation due to the flattening of the PQ potential by the non-minimal coupling.
    \item Amplification of scalar perturbations near the PQSB epoch, sourcing second-order gravitational waves (GW) potentially detectable by next-generation space-based experiments.
\end{enumerate}
Furthermore, our framework adheres to general constraints on axion mass and decay constant, providing a broad perspective without adhering to specific axion models or production mechanisms. Although for demonstration we will use the Starobinsky model, our approach works for generic inflaton scalar potentials.

{\bf Background and motivation}. The origin of the axion is described by the Peccei--Quinn theory, where a complex scalar with an appropriate potential spontaneously breaks the PQ symmetry $U(1)_{\rm PQ}$ at some high energy scale. Aside from the QCD axion, other axions and axion-like particles are expected to arise in string theory due to topological properties of extra dimensions \cite{Arvanitaki:2009fg}. These axions and the respective Peccei--Quinn symmetries can play important roles in the evolution of the universe, for example affecting inflation, reheating, and dark matter abundance.

When it comes to the QCD axion, its mass can be constrained from combined astrophysical and QCD effects as \cite{Chadha-Day:2021szb}
\begin{equation}\label{axion_mass_constr}
    10^{-11}\,{\rm eV}\lesssim m_a\lesssim 10^{-2}\,{\rm eV}~,
\end{equation}
which translates into constraints on the decay constant,
\begin{equation}\label{axion_f_constr}
    10^{9}\,{\rm GeV}\lesssim f\lesssim 10^{17}\,{\rm GeV}~,
\end{equation}
by using the one-to-one relation \cite{GrillidiCortona:2015jxo}
\begin{equation}\label{axion_m_vs_f}
    m_a=5.7\times 10^{-6}\,{\rm eV}\left(\frac{10^{12}\,{\rm GeV}}{f}\right)~.
\end{equation}
Tighter constraints can be obtained when considering specific models and production mechanisms of axion dark matter. We will use the constraints \eqref{axion_mass_constr} and \eqref{axion_f_constr} as guidelines, without referring to a particular axion model or production mechanism, to keep the framework as general as possible.

In the context of inflationary cosmology, PQSB is usually assumed to occur either before observable inflation (pre-inflationary PQSB) or after inflation (post-inflationary PQSB). In the former case, cosmic strings that form during the PQ phase transition are diluted by inflation, and the misalignment mechanism is regarded as the main source of axion dark matter abundance. This scenario, however, can be problematic due to the presence of the (massless) axion during the horizon exit of the CMB scales. This presence of axion can potentially lead to large isocurvature perturbations. In the post-inflationary PQSB, the cosmic strings, and subsequently domain walls, form after inflation, and can be dangerous because they can overclose the universe. The decay of these cosmic defects also contributes to the axion dark matter abundance in this case, constraining the models. Therefore, it is important to study possible PQ dynamics during and after inflation to understand axion dark matter.

Recently, in addition to pre- and post-inflationary PQSB, some attention has gone to a third scenario, where the PQ symmetry is broken in the middle of (observable) inflation, first proposed in \cite{Linde:1991km}. The authors of \cite{Redi:2022llj} and \cite{Harigaya:2022pjd} studied the production mechanisms and constraints on axion dark matter in this scenario, while in \cite{Ge:2023rce,Ge:2023rrq} the authors showed that the cosmic string-domain wall network can produce a large stochastic gravitational wave background, potentially relevant for pulsar timing array experiments, as well as primordial black holes (PBH) (see also \cite{Takahashi:2020tqv,Gonzalez:2022mcx,Kitajima:2023kzu} for axion domain walls and their cosmological signatures). A general proposal of how to achieve meso-inflationary PQSB is to consider a time-dependent (e.g. inflaton-dependent) PQ mass that can turn tachyonic at some point during inflation, triggering the symmetry breaking. A scenario where PQ phase transition happens during warm inflation was studied in \cite{Rosa:2021gbe}.

In our scenario, inflation is split into two stages: 1) The first stage is driven by a (PQ-neutral) inflaton field $\phi$, while the PQ symmetry is restored by the non-minimal coupling (CMB scales exit the horizon at this stage). 2) The second stage starts after the critical point in the potential is reached, and the PQ scalar $\rho$ acquires a tachyonic mass. Around the critical point, the classical force on the $\rho$-field gives way to quantum diffusion, which kicks $\rho$ away from the origin. This triggers the PQSB and starts the second slow-roll stage, which is made possible thanks to the flattening of the PQ potential by the non-minimal coupling. The difference with the usual Higgs/Peccei--Quinn inflationary models is that in our case, the PQ-driven inflation happens around the symmetric point of the potential rather than at large field values. For the inflaton sector, we can, in principle, use any single-field model, but as a benchmark example, we will use the Starobinsky model. This scenario can be described as hybrid inflation \cite{Linde:1993cn,Garcia-Bellido:1996mdl,Garcia-Bellido:1996oso} with a long waterfall regime, where the waterfall field is identified with the PQ radial scalar.

The paper is organized as follows. Section \ref{Sec_PQ} studies the case of single-field Peccei--Quinn inflation around the symmetric point and demonstrate the effect of the non-minimal coupling on the duration of inflation. Section \ref{Sec_two-field} brings together the inflaton and non-minimally coupled Peccei--Quinn Lagrangians, and then studies the resulting two-field inflation, both with analytical estimates and numerically. At this point we introduce the method of stochastic inflation to study the transient stage where the symmetry breaking happens. Section \ref{Sec_perturbations} focuses on scalar perturbations and shows how the power spectrum is amplified at small scales. Section \ref{Sec_discussion} concludes the paper and discusses about future works.

\section{Peccei--Quinn inflation around the symmetric point}\label{Sec_PQ}

Before studying the two-field dynamics of the inflaton $\phi$ and PQ scalar $\rho$, let us consider whether the simplest PQ potential is suitable for inflation around its symmetric point.

\subsection{Minimal coupling}

The simplest PQ Lagrangian is given by
\begin{equation}\label{L_PQ_min}
    \sqrt{-g}^{-1}\cl=\tfrac{1}{2}R-\partial S\partial\overbar S-\tfrac{1}{4}\lambda(f^2-2S\overbar S)^2~,
\end{equation}
where $S$ is the PQ complex scalar, which can be parametrized as $S=\frac{1}{\sqrt{2}}\rho e^{i\sigma}$. The angular mode $\sigma$ shifts by a real constant under the $U(1)_{\rm PQ}$ rotation and is identified with the QCD axion. The radial part $\rho$ eventually acquires non-vanishing VEV $\langle \rho\rangle=f$, spontaneously breaking the PQ symmetry.

As usual, to study inflation, we take the FLRW metric $ds^2=-dt^2+a^2(t)\delta_{ij}dx^idx^j$ and homogeneous background field $\rho(t)$, and obtain from Einstein equations,
\begin{equation}
    3H^2=\tfrac{1}{2}\dot\rho^2+V~,~~~\dot H=-\tfrac{1}{2}\dot\rho^2~,
\end{equation}
where $H\equiv\dot a/a$ is the Hubble function.
The scalar field (background) equation of motion (EOM) is
\begin{equation}\label{min_PQ_KG}
    \ddot\rho+3H\dot\rho+V_{,\rho}=0~,
\end{equation}
where subscript $\rho$ denotes the respective partial derivative. We define the potential and Hubble slow-roll parameters as
\begin{equation}
    \epsilon_V\equiv\frac{V^2_{,\rho}}{2V^2}~,~~~\eta_V\equiv\frac{V_{,\rho\rho}}{V},~~~\epsilon_H\equiv -\frac{\dot H}{H^2}~,~~~\eta_H\equiv\frac{\dot\epsilon_H}{H\epsilon_H}~.
\end{equation}
The usual slow-roll inflation is characterized by the conditions $\epsilon_V,\epsilon_H,|\eta_V|,|\eta_H|\ll 1$, under which the potential and Hubble slow-roll parameters are related to each other as
\begin{equation}
    \epsilon_V\simeq\epsilon_H,~~~\eta_V\simeq 2\epsilon_H-\tfrac{1}{2}\eta_H~.
\end{equation}

Inflation can also happen under ultra-slow-roll (USR) regime, where the slope of the potential in Eq. \eqref{min_PQ_KG} nearly vanishes, rendering the first two terms proportional to each-other. In this case, by writing $\eta_H=2\ddot\rho/(H\dot\rho)+2\epsilon_H$ (where we used $\epsilon_H=\dot\rho^2/(2H^2)$), and using $\ddot\rho=-3H\dot\rho$, we get $\eta_H\approx -6$ assuming that $\epsilon_H$ is small. Since during ultra-slow-roll $V_{,\rho}$ is much smaller than $3H\dot\rho$, we have the hierarchy $\epsilon_V\ll\epsilon_H$. This can lead to overestimation of the number of efolds if in the equation
\begin{equation}\label{N_formula}
    \Delta N=\int_{\rho_1}^{\rho_2} dt H=\int_{\rho_1}^{\rho_2} d\rho H/\dot\rho=\int_{\rho_1}^{\rho_2}\frac{d\rho}{\sqrt{2\epsilon_H}}~,
\end{equation}
we replace $\epsilon_H$ with $\epsilon_V$ in order to analytically estimate $\Delta N$. Nevertheless, analytical estimates based on $\epsilon_V$ can still be useful for obtaining an upper bound
\begin{equation}\label{efold_upper_bound}
    \Delta N\lesssim\int_{\rho_1}^{\rho_2}\frac{d\rho}{\sqrt{2\epsilon_V}}~.
\end{equation}
If regular slow-roll holds between $\rho_1$ and $\rho_2$, this upper bound is saturated. Since a sufficiently small initial value of $\rho$ can lead to vanishing $V_{,\rho}$, USR is possible in the case of PQ inflation around the origin of $\rho$. We note that the parameters of the PQ potential in our scenario are not necessarily fixed by the CMB observations because the CMB scales exit the horizon during the $\phi$-driven inflation, when $\rho$ is assumed to be stabilized around zero. Rather, we can expect a certain combination of the inflation and PQ parameters to be fixed by CMB data.

Let us now analytically estimate the duration of inflation in the PQ model \eqref{L_PQ_min}, which yields the following potential slow-roll parameters,
\begin{equation}\label{V_SR_PQ_min}
    \epsilon_V=\frac{8\rho^2}{(f^2-\rho^2)^2}~,~~~\eta_V=-4\frac{f^2-3\rho^2}{(f^2-\rho^2)^2}~.
\end{equation}
We immediately see that when $\rho\ll f$ and for $f\ll 1$ (in Planck units), $\eta_V$ becomes large, while $\epsilon_V$ can still be small if the initial value of $\rho$ is small enough. We can use $\epsilon_V$ from \eqref{V_SR_PQ_min} to estimate the upper bound on the number of efolds \eqref{efold_upper_bound}. To get the conservative upper bound we can take the final value of $\rho$ at its VEV, $\rho_2=f$. The resulting $\Delta N$ is
\begin{equation}\label{N_estimate_PQ}
    \Delta N\lesssim \frac{f^2}{4}\big(\log\frac{f}{\rho_1}-\frac{1}{2}\big)~.
\end{equation}
Here $\rho_1$ is the initial value of $\rho$, which is small, $\rho_1\ll f$. The estimate \eqref{N_estimate_PQ} tells us that the duration of inflation in this model is severely suppressed if we assume $f\ll 1$. More concretely, if we take for example a (an optimistically) small value $\rho_1/f=10^{-10}$, we get $\Delta N\lesssim 6f^2$ (we also confirm these estimates by numerically solving the EOM).

Remarkably, we find that a simple non-minimal coupling of the PQ field to scalar curvature ($\sim\rho^2 R$) can dramatically increase the number of efolds of the Peccei--Quinn inflation, and at the same time restore $U(1)_{\rm PQ}$ during the $\phi$-driven first stage of inflation, without adding explicit $\phi-\rho$ interactions to the Lagrangian.

\subsection{Non-minimal coupling}\label{sec_NM_PQ}

The Lagrangian for the non-minimally coupled PQ field is
\begin{equation}\label{L_PQ_nm}
    \sqrt{-g}^{-1}\cl=\tfrac{1}{2}A(S\overbar S)R-\partial S\partial\overbar S-\tfrac{\lambda}{4}(f^2-2S\overbar S)^2~.
\end{equation}
By rescaling the metric as
\begin{equation}
    g_{mn}\rightarrow A^{-1}g_{mn},
\end{equation}
we can bring the action to the Einstein frame,
\begin{equation}
    \sqrt{-g}^{-1}\cl=\tfrac{1}{2}R-A^{-1}\partial S\partial\overbar S-\tfrac{3}{4}A^{-2}\partial A\partial A-\tfrac{\lambda}{4}A^{-2}(f^2-2S\overbar S)~.
\end{equation}
As usual, we take a linear (in $S\overbar S$) form of the non-minimal coupling function,
\begin{equation}
    A=\mu^2+2\xi S\overbar S=\mu^2+\xi\rho^2~,
\end{equation}
where $\mu$ is a real parameter which can be eliminated by requiring that $A=M_P^2$ at the vacuum, or in Planck units,
\begin{equation}
    \langle A\rangle=\mu^2+\xi f^2=1~,
\end{equation}
so that $\mu^2=1-\xi f^2$. Then we can write
\begin{equation}\label{A_form}
    A=1-\xi(f^2-\rho^2)~,
\end{equation}
and the Lagrangian becomes (ignoring the axion),
\begin{equation}\label{L_NM_rho}
    \sqrt{-g}^{-1}\cl=\tfrac{1}{2}R-\tfrac{1}{2}G_{\rho\rho}\partial\rho\partial\rho-\frac{\lambda(f^2-\rho^2)^2}{4[1-\xi(f^2-\rho^2)]^2}~,
\end{equation}
where the field-space metric of $\rho$, denoted $G_{\rho\rho}$, is
\begin{equation}
    G_{\rho\rho}=\frac{1-\xi f^2+(1+6\xi)\xi\rho^2}{[1-\xi(f^2-\rho^2)]^2}~.
\end{equation}
The canonical parametrization of $\rho$ is given by the solution to
\begin{equation}\label{rho_tilde_eq}
    \frac{d\tilde\rho}{d\rho}=\sqrt{G_{\rho\rho}}~,
\end{equation}
where $\tilde\rho$ is the canonical scalar.

The second derivative of the scalar potential of Eq. \eqref{L_NM_rho} at $\rho=0$ is $V_{,\rho\rho}\simeq -\lambda f^2/(1-\xi f^2)^3$. This shows that large negative $\xi f^2$ flattens the potential while keeping the negative sign of $V_{,\rho\rho}$ and the correct sign of the kinetic term of $\rho$ around $\rho=0$.

Having taken into account the canonical parametrization \eqref{rho_tilde_eq}, the potential slow-roll parameters become
\begin{equation}\label{eps_V_PQnm}
    \epsilon_V=\frac{V_{,\tilde\rho}^2}{2V^2}=\frac{V^2_{,\rho}}{2G_{\rho\rho}V^2}=\frac{8\rho^2}{(f^2-\rho^2)^2(A+6\xi^2\rho^2)}\simeq \frac{8\rho^2}{f^4(1-\xi f^2+6\xi^2\rho^2)}~.
\end{equation}
and
\begin{equation}\label{eta_V_PQnm}
    \eta_V=\frac{V_{,\tilde\rho\tilde\rho}}{V}=\frac{V_{,\rho\rho}}{G_{\rho\rho}V}-\frac{\partial_\rho G_{\rho\rho}V_{,\rho}}{2G^2_{\rho\rho}V}\simeq -\frac{4(1-\xi f^2)^2}{f^2(1-\xi f^2+6\xi^2\rho^2)^2}~,
\end{equation}
where the approximation $\rho\ll f$ simplifies the expressions, and $A$ is given by \eqref{A_form}. 

For $|\xi|f^2\gg 1$, $\epsilon_V$ is suppressed compared to the minimally coupled PQ model. For $\eta_V$ in Eq. \eqref{eta_V_PQnm} the situation is more interesting. At $\rho=0$ (and when $6|\xi|\rho^2\ll f^2$), it reduces to $\eta_V=-4/f^2$, which is the same value as in the minimally coupled PQ model. Nonetheless, as $\rho$ moves away from the origin, it can be quickly suppressed once $6\xi^2\rho^2$ becomes much larger than $|\xi|f^2$ (this can happen while $\rho/f$ is still small because we assume $|\xi|\gg f^{-2}\gg 1$). Therefore, we can anticipate a larger $\Delta N$ in this case if the non-minimal coupling is sufficiently large. Indeed, we find that the number of efolds between some small $\rho_1$ and $\rho_2=f$ is
\begin{equation}\label{N_estimate_NM_PQ}
    \Delta N\simeq\int_{\tilde\rho_1}^{\tilde\rho_2}\frac{d\tilde\rho}{\sqrt{2\epsilon_V}}=\int_{\rho_1}^{\rho_2}d\rho\sqrt{\frac{G_{\rho\rho}}{2\epsilon_V}}\simeq \frac{f^2}{4}\big(\log\frac{f}{\rho_1}-\frac{1}{2}\big)+\frac{3}{4}\Big[|\xi|f^2-\log(1+|\xi|f^2)\Big]~,
\end{equation}
where we used $\rho_1\ll f$ and $\xi=-|\xi|$. As can be seen, the number of efolds is roughly proportional to $|\xi|f^2$ for large $|\xi|f^2$. This opens up a possibility of meso-inflationary PQ symmetry breaking, where the symmetry is broken for $\Delta N\gg 1$ efolds before the end of inflation.

\section{Two-field inflation with Peccei--Quinn phase}\label{Sec_two-field}

We now consider effective two-field inflation where the usual inflaton scalar $\phi$ is responsible for the first $\Delta N_1$ efolds of inflation, while the non-minimally coupled PQ field $\rho$ (either alone or in combination with $\phi$) drives the remaining $\Delta N_2$ efolds, such that $\Delta N_1+\Delta N_2=50\sim 60$. The Lagrangian reads
\begin{equation}\label{L_nm}
    \sqrt{-g}^{-1}\cl=\tfrac{1}{2}A(S\overbar S)R-\tfrac{1}{2}\partial\phi\partial\phi-\partial S\partial\overbar S-U_1(\phi)-U_2(S\overbar S)~,
\end{equation}
where $U_1$ and $U_2$ are an inflaton potential and a PQ potential, respectively (we denote Jordan-frame potentials by $U$, and Einstein-frame potentials by $V$). As a concrete example we take the Starobinsky potential \cite{Starobinsky:1980te} for the inflaton, and the usual sombrero-shaped potential for the PQ field,~\footnote{A non-minimally coupled PQ model where both $\rho$ and $\sigma$ participate during inflation was studied in \cite{McDonough:2020gmn,Lorenzoni:2024krn}.}
\begin{align}
    U_1 &=\tfrac{3}{4}M^2\big(1-e^{-\sqrt{\frac{2}{3}}\phi}\big)^2~,\label{U_1}\\
    U_2 &=\tfrac{\lambda}{4}(f^2-2S\overbar S)^2~.\label{U_2}
\end{align}
We do not consider any direct couplings between $\phi$ and $S$ in the Jordan frame. If $\phi-S$ coupling exists in the Jordan frame, it can affect the inflationary history depending on the form of the coupling. For example, it is possible to restore the PQ symmetry during early inflation without the non-minimal coupling to gravity (or with a positive sign of the parameter $\xi$ in \eqref{A_form2}), by introducing a suitable $\phi$-dependent effective mass for $\rho$ in the Lagrangian. However, the simplest ``sombrero" potential is not flat enought to realize a $\rho$-driven inflationary phase for subplanckian decay constant $f$. This would require further modification of the scalar potential. The non-minimal coupling of $\rho$ can naturally achieve both the restoration of $U(1)_{\rm PQ}$ during early inflation, and flattening of the potential in $\rho$-direction. Then, for simplicity we assume that the $\phi-S$ couplings in the Jordan frame are absent or suppressed.

After the Weyl rescaling, from \eqref{L_nm} we obtain the Einstein frame Lagrangian,
\begin{equation}
    \sqrt{-g}^{-1}\cl=\tfrac{1}{2}R-\tfrac{1}{2}A^{-1}\partial\phi\partial\phi-\tfrac{3}{4}A^{-2}\partial A\partial A-A^{-1}\partial S\partial\overbar S-A^{-2}(U_1+U_2)~,
\end{equation}
where $\phi-S$ interactions through the function $A(S\overbar S)$ are now generated, both in kinetic and potential terms.

As discussed in the previous section, we take
\begin{equation}\label{A_form2}
    A=1-\xi(f^2-\rho^2)~,
\end{equation}
such that at the vacuum, $A=1$. This leads to the effective two-field Lagrangian (writing in terms of $\rho=\sqrt{2}|S|$)
\begin{equation}\label{L_NM_rho_2}
    \sqrt{-g}^{-1}\cl=\tfrac{1}{2}R-\tfrac{1}{2}G_{\phi\phi}\partial\phi\partial\phi-\tfrac{1}{2}G_{\rho\rho}\partial\rho\partial\rho-A^{-2}(U_1+U_2)~,
\end{equation}
where the field-space metric reads
\begin{equation}\label{G_two_field}
    G_{\phi\phi}=\frac{1}{1-\xi(f^2-\rho^2)}~,~~~G_{\rho\rho}=\frac{1-\xi f^2+(1+6\xi)\xi\rho^2}{[1-\xi(f^2-\rho^2)]^2}~.
\end{equation}

Klein--Gordon equations with non-trivial field-space metric can be written with the help of covariant formalism,
\begin{equation}
    \Box\Phi^C+\Gamma^C_{AB}\partial\Phi^A\partial\Phi^B-G^{AC}\partial_AV=0~,
\end{equation}
where the indices $A,B,C$ denote (real) scalar fields of the model, $G^{AB}$ is the inverse of the field-space metric, $\Gamma^C_{AB}$ are the corresponding Christoffel symbols, and $\Box=\nabla_\mu\nabla^\mu$ ($\partial$ without explicit indices denote space-time derivatives). For the Lagrangian \eqref{L_NM_rho_2} with the field-space metric \eqref{G_two_field}, we get the background equations
\begin{align}
    \ddot\phi+3H\dot\phi+G^{\phi\phi}(\partial_\rho G_{\phi\phi}\,\dot\phi\dot\rho+V_{,\phi}) &=0~,\label{phi_EOM_full}\\
    \ddot\rho+3H\dot\rho+\tfrac{1}{2}G^{\rho\rho}(\partial_\rho G_{\rho\rho}\dot\rho^2-\partial_\rho G_{\phi\phi}\dot\phi^2+2V_{,\rho}) &=0~,\label{rho_EOM_full}
\end{align}
and Friedmann equations,
\begin{align}
    3H^2 &=\tfrac{1}{2}(G_{\phi\phi}\dot\phi^2+G_{\rho\rho}\dot\rho^2)+V~,\\
    \dot H &=-\tfrac{1}{2}(G_{\phi\phi}\dot\phi^2+G_{\rho\rho}\dot\rho^2)~.
\end{align}

For future convenience, we write the Einstein-frame scalar potential $V=A^{-2}(U_1+U_2)$ as
\begin{equation}\label{V_final}
    V=\frac{3M^2(1-x)^2+\lambda(f^2-\rho^2)^2}{4[1-\xi(f^2-\rho^2)]^2}=3M^2\frac{(1-x)^2+\hat\lambda(1-\hat\rho^2)^2}{4[1+\hat\xi(1-\hat\rho^2)]^2}~,
\end{equation}
where $x\equiv e^{-\sqrt{2/3}\phi}$, $\xi<0$ (as assumed earlier), and
\begin{equation}
    \hat\rho\equiv \rho/f~,~~~\hat\xi\equiv |\xi|f^2~,~~~\hat\lambda\equiv \frac{\lambda f^4}{3M^2}~,
\end{equation}
such that $\langle\hat\rho\rangle=1$ (not to confuse with $\tilde\rho$ which we define as the canonical PQ scalar). The parameter $\hat\xi$ controls the flatness of the PQ potential, and $\hat\lambda$ shows the magnitude of the PQ potential relative to the Starobinsky potential (during early inflation when $x$ and $\hat\rho$ are small).

\subsection{First stage}

We define the first stage of inflation as starting from the horizon exit of CMB scales, when $\rho$ is stabilized around zero and $\phi\gg 1$ (or $x\ll 1$), and ending when the effective mass of the $\rho$-field vanishes, at some critical value $\phi_c$ of the inflaton. After $\phi_c$, the mass-squared of $\rho$ becomes negative, and quantum fluctuations destabilize the classical trajectory from $\rho=0$, and the second stage begins. During the transition, slow roll may or may not be violated, depending on whether or not $\epsilon_H$ becomes large, $\epsilon_H\geq 1$, before we reach the critical point. From the potential \eqref{V_final}, by setting $V_{,\rho\rho}|_{\rho=0}=0$, we find the critical value,
\begin{equation}\label{x_c}
    x_c\equiv e^{-\sqrt{\frac{2}{3}}\phi_c}=1-\sqrt{\hat\lambda/\hat\xi}~.
\end{equation}
Nevertheless, if the velocity of $\phi$ becomes non-negligible near $\phi_c$, the effective (canonical) mass of $\rho$ around $\rho=0$ gains the corresponding correction term thanks to the non-trivial field-space metric (see Eq. \eqref{rho_EOM_full}),
\begin{align}
\begin{aligned}
    m_{\tilde\rho,{\rm eff}}^2 &=\partial_\rho(G^{\rho\rho}V_{,\rho}-\tfrac{1}{2}G^{\rho\rho}\partial_\rho G_{\phi\phi}\dot\phi^2)|_{\rho=0}\\
    &=\frac{3M^2\hat\xi}{f^2A_0^2}\Big[(1-x)^2-\frac{\hat\lambda}{\hat\xi}-\frac{\epsilon_\phi[(1-x)^2+\hat\lambda]}{2(3-\epsilon_\phi)}\Big]~,\label{m2_eff_full}
\end{aligned}
\end{align}
where $\epsilon_\phi\equiv\epsilon_H|_{\dot\rho=0}=G_{\phi\phi}\dot\phi^2/(2H^2)$ as the $\phi$-driven slow-roll parameter relevant for the first stage of inflation, $G_{\phi\phi}=G_{\rho\rho}=1/A_0$ (with $A_0=1+\hat\xi$), and $H^2=V/(3-\epsilon_\phi)$ at $\rho=\dot\rho=0$. The effective critical point in time, $t_c^{\rm eff}$, can be found by solving $m^2_{\tilde\rho,{\rm eff}}=0$, and the effective critical value of the inflaton is then given by $\phi^{\rm eff}_c=\phi(t_c^{\rm eff})$, which approaches $\phi_c$ from \eqref{x_c} as $\epsilon_\phi\rightarrow 0$.

During early inflation, when $x,\epsilon_\phi\rightarrow 0$, we have
\begin{equation}\label{rho_mass_I}
    m^2_{\tilde\rho,{\rm eff}}\simeq\frac{3M^2}{f^2A_0^2}(\hat\xi-\hat\lambda)~,
\end{equation}
and for the stability of $\rho$ in this regime (and also for the existence of the critical point \eqref{x_c}), we require $\hat\xi>\hat\lambda$. As $\phi$ approaches $\phi^{\rm eff}_c$, the effective mass-squared \eqref{m2_eff_full} decreases, and once it vanishes, quantum diffusion in $\rho$-direction becomes important. We will discuss this regime in the next section and focus on the first stage here.

During early first stage (at $\rho,x\rightarrow 0$), the scale of inflation is given by the approximate potential
\begin{equation}\label{V_sI_model_1}
    V\simeq\frac{3M^2}{4A_0^2}(1+\hat\lambda)~,
\end{equation}
where we can identify three different scenarios depending on the hierarchy of the parameters,
\begin{equation}
    {\rm (a)}~~\hat\lambda\sim 1~,~~~{\rm (b)}~~\hat\lambda\ll 1~,~~~{\rm (c)}~~\hat\lambda\gg 1~.\label{case_abc}
\end{equation}
In case (a), both the Starobinsky potential and PQ potential are comparable in magnitude; in case (b), the first stage is dominated by the Starobinsky potential, $V\sim\frac{3}{4}M^2/A^2_0$; in case (c), it is dominated by the constant term of the PQ potential, $V\sim\frac{1}{4}\lambda f^4/A_0^2$. In each case, the CMB scale amplitude of scalar perturbations \cite{Planck:2018jri},
\begin{equation}\label{Amp_CMB}
    P_\zeta\simeq\frac{V}{24\pi^2\epsilon_V}\approx 2.1\times 10^{-9}~,
\end{equation}
can be used to fix one relevant parameter, or a combination of parameters. From the potential slow-roll parameters (at $\rho=\dot\rho=0$) at the horizon exit, we can obtain the scalar spectral index and tensor-to-scalar ratio,
\begin{equation}\label{nsr_eqs}
    n_s\simeq 1+2\eta_V-6\epsilon_V~,~~~r\simeq 16\epsilon_V~,
\end{equation}
which are constrained by observations as \cite{Planck:2018jri}
\begin{equation}
    n_s\approx 0.9649\pm 0.0042~(68\%{~\rm CL})~,~~~r<0.064~(95\%{~\rm CL})~.
\end{equation}

Note that the slow-roll parameters $\epsilon_V$ and $\eta_V$ during the first stage are modified compared to the regular Starobinsky inflation due to the non-canonical kinetic term of the inflaton $\phi$. In particular, when $\rho=0$, the coefficient of the kinetic term is $G_{\phi\phi}|_{\rho=0}=1/A_0=1/(1+\hat\xi)$. This means that the appropriate slow-roll parameters are given by
\begin{align}
    \epsilon_V &=\frac{A_0V^2_{,\phi}}{2V^2}=\frac{4A_0 x^2(1-x)^2}{3[\hat\lambda+(1-x)^2]^2}~,\label{epsilon_V_phi}\\
    \eta_V &=\frac{A_0V_{,\phi\phi}}{V}=-\frac{4A_0x(1-2x)}{3[\hat\lambda+(1-x)^2]}~,\label{eta_V_phi}
\end{align}
and the scalar power spectrum during the first stage is
\begin{equation}\label{P_I_general}
    P_\zeta\simeq\frac{3M^2[\hat\lambda+(1-x_*)^2]^3}{128A_0^3\pi^2x_*^2(1-x_*)^2}~,
\end{equation}
where $x_*$ is the CMB value of $x$.

\begin{figure}
\centering
  \includegraphics[width=.95\linewidth]{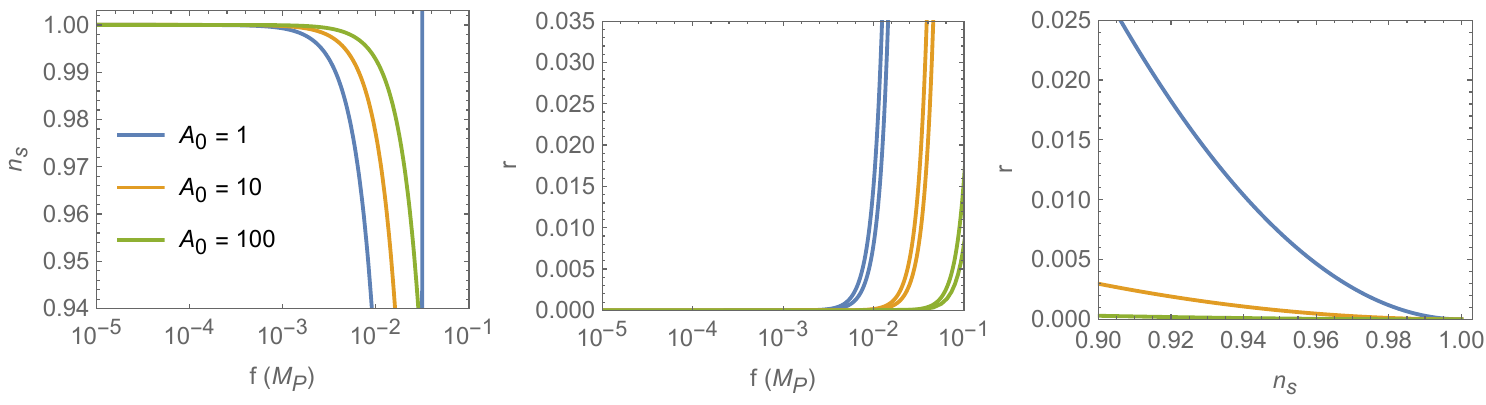}
\captionsetup{width=1\linewidth}
\caption{Parametric plots of scalar tilt $n_s(x_*)$, tensor-to-scalar ratio $r(x_*)$, and the axion decay constant $f(x_*)$ in the Starobinsky--Peccei--Quinn hybrid inflation. Plots cover the whole range $0<x_*<1$ ($x_*\equiv e^{-\sqrt{2/3}\phi_*}$ corresponds to the horizon exit of the CMB scale) for $A_0=1,10,100$ and $\hat\lambda=1$. The vertical blue line in the left plot represents the values of $\phi_*$ close to $0.6$, leading to large $r$ of around $0.9$, and can be ignored (analogous orange and green vertical lines are outside of the plot, corresponding to $f\gtrsim 1$).}
\label{Fig_case_a_nsr}
\end{figure}

{\bf Case (a).} We begin with the case free of strong hierarchy between the parameters $3M^2$ and $\lambda f^4$, i.e. where $\hat\lambda$ is close to $\co(1)$. Using the CMB value $P_\zeta\approx 2.1\times 10^{-9}$, from \eqref{P_I_general} we get
\begin{equation}
    \frac{M^2}{A_0^3}\simeq 8.84\times 10^{-7}\frac{x_*^2(1-x_*)^2}{[\hat\lambda+(1-x_*)^2]^3}~.
\end{equation}
By combining the above expression and the definition $\hat\lambda\equiv\lambda f^4/(3M^2)$, we can write
\begin{equation}\label{f4_eq}
    f^4\simeq 2.65\times 10^{-6}\frac{\hat\lambda A_0^3 x_*^2(1-x_*)^2}{\lambda[\hat\lambda+(1-x_*)^2]^3}~.
\end{equation}
We can then relate permitted values of $n_s$ and $r$ to the axion decay constant $f$ by using their slow-roll expressions \eqref{nsr_eqs},
\begin{align}
    n_s &\simeq 1-\frac{8A_0x_*(1-2x_*)}{3[\hat\lambda+(1-x_*)^2]}-\frac{8A_0x_*^2(1-x_*)^2}{[\hat\lambda+(1-x_*)^2]^2}~,\label{n_s_xstar}\\
    r &\simeq \frac{64A_0x_*^2(1-x_*)^2}{3[\hat\lambda+(1-x_*)^2]^2}~,\label{r_xstar}
\end{align}
and drawing parametric plots of $n_s(x_*)$ versus $f(x_*)$, and $r(x_*)$ versus $f(x_*)$ for the range $0<x_*<1$. See Fig. \ref{Fig_case_a_nsr} (left and center). Fig. \ref{Fig_case_a_nsr} (right) also plots possible values of $n_s(x_*)$ against $r(x_*)$. In these figures, we fix $\hat\lambda=1$, and take $A_0=1,10,100$ for instances ($A_0=1$ corresponds to $\hat\xi=0$, i.e. the minimal coupling limit). The main takeaway from these plots is that even without studying the second stage, we can conclude that $f$ is bounded from below as $\gtrsim 10^{-2}$ in case (a). In addition, from the $n_s-r$ plot of Fig. \ref{Fig_case_a_nsr}, we have learned that large $A_0$ (viz large $\hat\xi$) can reduce the tensor-to-scalar ratio significantly.

{\bf Case (b).} We proceed to the small $\hat\lambda$ case where the potential during the first stage can be approximated by that of the Starobinsky,
\begin{equation}\label{V_case_a}
    V\simeq\frac{3M^2}{4A_0^2}(1-x)^2~.
\end{equation}
In this case, from the amplitude \eqref{Amp_CMB}, we can obtain (at the leading order in $\hat\lambda$)
\begin{equation}\label{M_CMB}
    M\simeq 0.94\times 10^{-3}\frac{A_0^{3/2}x_*}{(1-x_*)^2}~.
\end{equation}
The number of efolds \eqref{N_formula} during the first stage, i.e. between $x_*$ and $x_c$ (the critical point) is
\begin{equation}
    \Delta N_1\simeq\frac{3}{4A_0}\Big(\frac{x_c-x_*}{x_cx_*}+\log\frac{x_*}{x_c}\Big)~.
\end{equation}
The Starobinsky inflation is of the large-field type, i.e. horizon exit of the CMB scales occurs when $\phi$ is large, or $x_*\ll 1$, in which case we can approximate the above equation as
\begin{equation}\label{dN_x_approx}
    \Delta N_1\simeq\frac{3}{4A_0x_*}~,
\end{equation}
where $x_c$ is ignored because it gives only subleading contributions. After substituting $x_*$ from \eqref{dN_x_approx} into \eqref{M_CMB}, we obtain
\begin{equation}
    M\approx 7.05\times 10^{-4}\frac{\sqrt{A_0}}{\Delta N_1}~.
\end{equation}
Similarly, for $n_s$ and $r$ we find
\begin{equation}
    n_s\simeq 1-\frac{2}{\Delta N_1}~,~~~r\simeq \frac{12}{A_0\Delta N_1^2}~,
\end{equation}
by using \eqref{n_s_xstar} and \eqref{r_xstar}. The difference from the predictions of the usual Starobinsky model can be seen in the expression for $r$ containing $A_0=1+\hat\xi$, which comes from the non-canonical kinetic term of $\phi$. For large $A_0$, tensor-to-scalar ratio is suppressed, in line with our results in Fig. \ref{Fig_case_a_nsr}. If the second stage is long, e.g. $\co(10)$ efolds, $\Delta N_1$ will be reduced from the usual $50\sim 60$ efolds, and the value of $n_s$ will also decrease from standard prediction of the Starobinsky inflation. This can be addressed by modifying the Starobisnky potential~\footnote{For example this can be done by adding a negative power of $x$ to the potential with a small coefficient, in order to flatten the potential near the horizon exit of CMB scale, see e.g. Fig. 5 of \cite{Aldabergenov:2023yrk}.}, or by considering a different potential that predicts a larger $n_s$. One such example is the axion monodromy-type model with $V\sim \big[(1+\phi^2)^{1/4}-1\big]$. Using this as the Jordan frame potential \eqref{U_1}, we get approximately $n_s\simeq 1-5/(4\Delta N_1)$. Nevertheless, we do not know \textit{a priori} the value of $\Delta N_1$, and so the best-fitting inflaton potential should be chosen depending on the results from the second stage of inflation, which will depend on the parameter choice. Modifying the potential will affect the results quantitatively, but the qualitative behavior of the solutions are expected to hold, at least for large field inflation. For instance, the initial conditions for the $\rho$-driven second stage of inflation (derived from quantum diffusion near the symmetric point) can vary depending on the form of the inflaton potential. Yet, the slow-roll solution of the second stage is largely insensitive to these initial conditions due to the Hubble friction in Eq. \eqref{rho_EOM_full}. For concreteness, we will continue with the Starobinsky potential as a reference model, which will also allow us to use numerical methods when necessary.

We define the end of the first stage as the time when $x=x_c^{\rm eff}$, viz. when the effective $\rho$-mass vanishes, and the slow-roll regime may or may not hold at this point depending on the parameter choice. We can consider the expression
\begin{equation}
    m^2_{\tilde\rho,{\rm eff}}\propto \frac{(1-x)^2-\hat\lambda/\hat\xi}{(1-x)^2+\hat\lambda}-\frac{\epsilon_\phi}{2(3-\epsilon_\phi)}~,\label{m2eff_eq}
\end{equation}
obtained from \eqref{m2_eff_full} by dividing the expression in the square brackets by $(1-x)^2+\hat\lambda$. Upon slow-roll violation, e.g. at $\epsilon_\phi=1$, we check the sign of \eqref{m2eff_eq}: If it is negative (positive), the critical point is reached before (after) slow-roll violation. In case (b) ($\hat\lambda\ll 1$), if $\hat\lambda,\hat\lambda/\hat\xi\ll (1-x)^2$, the mass-squared is still positive at $\epsilon_\phi=1$, so slow-roll is broken before the critical point. If $\hat\lambda/\hat\xi$ is close to $(1-x)^2$, slow-roll may hold at $t_c^{\rm eff}$.

{\bf Case (c).} In this case, $\hat\lambda\gg 1$, i.e. the potential is dominated by the PQ constant term $\sim\lambda f^4$, which suppresses the slow-roll parameters in comparison to case (b), as is clear from \eqref{epsilon_V_phi} and \eqref{eta_V_phi}. The positivity of the effective $\rho$ mass-squared \eqref{rho_mass_I} during the first stage requires $\hat\xi>\hat\lambda$, causing $A_0\sim\hat\xi\gtrsim\hat\lambda$ for $\hat\lambda\gg 1$, i.e., $A_0$ cannot be small. Hence, both potential slow-roll parameters can be small only if $x\ll 1$. 

From the power spectrum \eqref{Amp_CMB} at the CMB scale, we can obtain the condition (for $\hat\lambda\gg 1$)
\begin{equation}\label{Amp_case_c}
    \frac{(\lambda f^4)^3}{A_0^3M^4}\simeq 2.4\times 10^{-5}x_*^2(1-x_*)^2~,
\end{equation}
and the number of efolds becomes
\begin{equation}\label{dN_case_c}
    \Delta N_1\simeq\frac{3\hat\lambda}{4A_0}\bigg[\frac{x_c-x_*}{x_cx_*}+\log\frac{x_c(1-x_*)}{x_*(1-x_c)}\bigg]~.
\end{equation}
Equation \eqref{dN_case_c} can re-express $M$ in terms of $\Delta N_1$ (after using $\hat\lambda=\lambda f^4/(3M^2)$) and casts \eqref{Amp_case_c} as
\begin{equation}\label{f_eq_case_c}
    f\simeq \frac{0.035A_0^{1/4}}{\sqrt{\Delta N_1}\lambda^{1/4}}\sqrt{x_*(1-x_*)}\bigg[\frac{x_c-x_*}{x_cx_*}+\log\frac{x_c(1-x_*)}{x_*(1-x_c)}\bigg]^{\frac{1}{2}}~.
\end{equation}
For large $\hat\lambda$, we can approximate $n_s$ as
\begin{equation}
    n_s\simeq 1-\frac{8A_0}{\hat\lambda}x_*(1-2x_*)~,
\end{equation}
which, after eliminating $\hat\lambda$ by \eqref{dN_case_c}, can be written as
\begin{equation}
    x_*(1-2x_*)\bigg[\frac{x_c-x_*}{x_cx_*}+\log\frac{x_c(1-x_*)}{x_*(1-x_c)}\bigg]\simeq \frac{1-n_s}{2}\Delta N_1~.
\end{equation}
By combining this equation with \eqref{f_eq_case_c} and using $n_s\approx 0.9649$~\footnote{Of course, the exact prediction for $n_s$ may be slightly different from $0.9649$, so \eqref{f_estimate} should be treated as a rough estimate.}, we get
\begin{equation}\label{f_estimate}
    f\simeq 4.64\times 10^{-3}\frac{A_0^{1/4}}{\lambda^{1/4}}\sqrt{\frac{1-x_*}{1-2x_*}}~,
\end{equation}
where it becomes clear that $f$ is bounded from below,
\begin{equation}\label{f_b_lower_bound}
    f\gtrsim 4.64\times 10^{-3}\frac{A_0^{1/4}}{\lambda^{1/4}}~,
\end{equation}
assuming $x_*<1/2$, as otherwise we will get a blue-tilted spectrum. Consistency with the astrophysical upper bound \eqref{axion_f_constr} ($f\lesssim 0.1$ in Planck units) requires that
\begin{equation}
    A_0=1+\hat\xi\lesssim 10^4~,
\end{equation}
for $\lambda=0.1$ or so. On the other hand, $\hat\xi$ is necessarily positive, rendering $A_0\geq 1$. Thus, $f$ in \eqref{f_b_lower_bound} cannot be smaller than $\co(10^{-3})$ in this case (we do not consider large $\lambda$ because it would break perturbative regime).

\subsection{Transient stage: quantum diffusion
and classicalization}

At the end of stage one, there is a brief transient stage where quantum
effects become non-negligible for the radial part of the PQ field $\rho$.
To see this, we first recall that $\rho$-field's value is effectively zero during the first stage due to the $U(1)_{\RM{PQ}}$ symmetry restored by the non-minimal coupling. As the $\rho$-field's effective mass becomes tachyonic,
the $\rho$-field now sits at an unstable critical point, where the uncertainty principle dictates that quantum fluctuations will push it down to the stable minima at $\rho=f$. The usual classical equations of motion for the $\rho$-field are not sufficient to evolve the system to the subsequent stages. Instead, a framework including the effects of quantum backreaction is needed.

Let us make our statements more concrete. When the effective mass (and the slope of the potential) of the $\rho$-field vanishes at the end of stage one, the field enters into the USR regime\footnote{The USR regime applies to the $\rho$-field only. The $\phi$-field can still be in slow-roll for a wide range of parameters (see next section for example), such that the first two Hubble slow-roll parameters, which capture the combined effects of $\rho$ and $\phi$ on the change of $H$, can both still be small. This is in contrast to typical single-field USR models where $\eta_H\sim -6+O(\epsilon_H)$.}, where quantum diffusion becomes important. Consider first the classical EOM \eqref{rho_EOM_full} for the $\rho$-field. At the end of stage one, the classical force $\rho$-field feels from the potential vanishes while $\rho\sim 0$. Force terms including
the factors
$\partial_{\rho}G_{\rho\rho}=0$ and $\partial_{\rho}G_{\phi\phi}=0$
also vanish. Thus, the $\rho$-velocity is suppressed due to the Hubble friction
\begin{equation}
	\ddot{\rho}+3H\dot{\rho}=0\Rightarrow
	\dot{\rho}\sim e^{-3H t}\,.
	\label{}
\end{equation}
This suppression implies that classical
evolution by itself favors the position
in field space where $\rho\sim 0$,
possibly extending inflation for too long and producing curvature perturbations that are too large (after inflation ends in the $\phi$-direction). Aside from velocity suppression,
the common assumption (eg. in \cite{tada2023primordial,tada2023stochastic}) of neglecting $\ddot{\rho}$ may not be taken for granted, and in our case may not be valid, even around $\rho=0$. In particular, for our two-field model, the second (Hubble) slow-roll parameter $\eta_H$ is not close to $-6$ (viz, the typical single-field USR value) but instead
\begin{equation}
	\eta_H\sim 2\epsilon_H+O(\ddot{\phi}/(3H\dot{\phi}))
	\label{}
\end{equation}
when $\rho,\dot{\rho}\sim 0$. If standard slow-roll still holds in the $\phi$-direction, such that $\ddot{\phi}\ll 3H\dot{\phi}$, the second slow-roll parameter satisfies $|\eta_H|\ll 1$ and stays small. Nonetheless, slow-roll can be temporarily violated for a certain parameter region.

The non-standard evolution described above signals the need for a quantum description beyond the traditional
slow-roll attractor description.
In principle, the full description we need is quantum gravity
because quantum diffusion occurs
via backreaction of the small-scale modes
on the (locally) 
homogeneous background field,
which couples to geometry through
the Friedmann equations.
Fortunately,
we expect the homogeneous part of the $\rho$-field to deviate only slightly
from a classical state,
provided the transient stage is brief 
(this briefness will be verified later). So, there is no need for a full quantum description. When this is indeed the case, the effective description provided by stochastic inflation 
\cite{starobinsky1986stochastic,vennin2020stochastic} can be applied.

In the framework of stochastic inflation,
quantum diffusion occurs when small-scale modes collectively backreact on the
(locally) homogeneous part of the field
as they get stretched by
(quasi) de Sitter expansion.
The classical equations of motion
for the background field
are thus modified with
noise terms arising from the quantum
fluctuations of these small-scale modes.
The homogeneous field becomes stochastic by inheriting the statistics of the
quantum noise terms through Langevin-like equations.
The statements can formally be made rigorous
starting from canonical quantization
and then obtaining an effective phase-space
distribution
of $\rho$ that mimics some (but not all) of the features of its true quantum wavefunction.

For our purpose,
we wish to use the stochastic formalism
to obtain the classical values of
$\rho$ and its conjugate momentum
after the brief transient stage
of quantum diffusion.
These values serve as initial
conditions for the subsequent classical
evolution, which we call stage II.
Multi-stage evolution containing a brief
quantum diffusion stage in between two classical phases is common in
mutli-field models, especially
in hybrid inflation
(\cite{clesse2010hybrid,Clesse:2015wea,Kawasaki:2015ppx,kodama2011on,tada2023stochastic} for example).
But because the classicalization process
in general is not fully
understood, the transition from
a quantum diffusion
stage (probabilistic) to the subsequent classical stage (deterministic)
is difficult to implement, especially numerically.
In some cases, one manually selects a set of
classical values after the quantum diffusion phase based on order-of-magnitude
estimates, without having to worry about
stitching stochastic evolution
to classical deterministic evolution.
In other cases where quantum diffusion is treated
more carefully with the stochastic formalism,
one sometimes assumes an effectively
instantaneous classicalizaton time for
smoother numerical implementation but
runs into the danger of \textit{a priori} excluding
the parameter space with prolonged
quantum diffusion,
which could affect primordial black hole
abundance \cite{Biagetti2018primordial}
if the $\rho$-field dominates the inflaton
dynamics. In the following, we will use effective
equations of motion for the quantum diffusion stage that are derived rigorously with canonical quantization and the stochastic framework. We will not ignore any time-derivative (or slow-roll) terms by default, as our model does not follow all the usual slow-roll assumptions. Lastly, we will employ a classicalization criteria, which helps determine the classicalization time, allowing for a smooth transition to the subsequent classical evolution.

\subsubsection{Quantum diffusion}
To describe quantum diffusion in the phase-space of the $\rho$-field we introduce the momentum field (we start with physical time in the Lagrangian \eqref{L_NM_rho_2} and then use $\frac{d}{dt}=H\frac{d}{dN}$):
\begin{equation}
    \pi_\rho\equiv\frac{1}{a^3H^2}\frac{\partial\cl(t(N))}{\partial\rho_{,N}}=G_{\rho\rho}\rho_{,N}~,~~~\pi^\rho=G^{\rho\rho}\pi_\rho=\rho_{,N}~,
    \label{rescaled-momentum}
\end{equation}
where the subscript ``$N$" denotes the derivative w.r.t. the efold number, and $G^{\rho\rho}$ raises the index (field metric is diagonal) and leads to the contravariant momentum. By this definition, from the $\rho$ EOM \eqref{rho_EOM_full}, we obtain the equation system
\begin{align}
\begin{aligned}
    \rho_{,N} &=\pi^\rho~,\\
    \pi_{,N}^\rho &= -\big(3-\epsilon_H+\tfrac{1}{2}G^{\rho\rho}\partial_\rho G_{\rho\rho}\pi^{\rho}\big)\pi^{\rho}+G^{\rho\rho}\big(\tfrac{1}{2}\partial_\rho G_{\phi\phi}\phi_{,N}^2-V_{,\rho}/H^2\big)~.
\end{aligned}
\end{align}
By splitting $\rho$ and $\pi^\rho$ into IR and UV modes (see Appendix \ref{App_A}), we can obtain the stochastic equations
\begin{align}
\begin{aligned}
    \rho_{,N} &=\pi^\rho+\xi_\rho~,\\
    \pi_{,N}^\rho &= -\big(3-\epsilon_H+\tfrac{1}{2}G^{\rho\rho}\partial_\rho G_{\rho\rho}\pi^{\rho}\big)\pi^{\rho}+G^{\rho\rho}\big(\tfrac{1}{2}\partial_\rho G_{\phi\phi}\phi_{,N}^2-V_{,\rho}/H^2\big)+\xi_{\pi^\rho}~,\label{rho_pi_eqs_NL}
\end{aligned}
\end{align}
where $\rho$ and $\pi^{\rho}$ should be understood as IR modes, in which the equations are in general non-linear, while they have been linearized in the UV modes, which in turn generate the stochastic noise terms $\xi_\rho$ and $\xi_{\pi^\rho}$ defined as
\begin{align}
	\xi_\rho &\equiv -\int\frac{{\rm d}^3{\bf k}}{(2\pi)^{3/2}}\partial_NW\Big(\frac{k}{k_s}\Big)\big(e^{-i{\bf k}{\bf x}}\rho_{\bf k}\hat a_{\bf k}+{\rm h.c.}\big)~,\\
    \xi_{\pi^\rho} &\equiv -\int\frac{{\rm d}^3{\bf k}}{(2\pi)^{3/2}}\partial_NW\Big(\frac{k}{k_s}\Big)\big(e^{-i{\bf k}{\bf x}}\pi^{\rho}_{\bf k}\hat a_{\bf k}+{\rm h.c.}\big)~,
\end{align}
where $W(x)=\theta(x-1)$ is a step function
that selects out large-wavenumber modes. To avoid confusion, $\xi$ and $\hat\xi$ without any indices always denote the non-minimal coupling, while $\xi$ with subscript $\rho$ or $\pi^\rho$ refers to the stochastic noise.

Near the origin $\rho=0$, we can further linearize Eqs. \eqref{rho_pi_eqs_NL} w.r.t. $\rho$ and $\pi^\rho$, but keep non-linearities in $\phi$ and its derivative, treating them as a classical dynamical background (possibly non-slow-roll when $\dot\phi$ is large),
\begin{align}
\begin{aligned}
    \rho_{,N} &=\pi^\rho+\xi_\rho~,\\
    \pi_{,N}^\rho &= -(3-\epsilon_\phi)\pi^{\rho}-H^{-2}m^2_{\tilde\rho,{\rm eff}}\,\rho+\xi_{\pi^\rho}~,\label{rho_pi_eqs_L}
\end{aligned}
\end{align}
where $G_{\rho\rho}\simeq G_{\phi\phi}\simeq 1/A_0$ at the leading order in $\rho$, and $\epsilon_\phi$ is defined as
\begin{equation}
    \epsilon_\phi\equiv\epsilon_H|_{\rho_{,N}=0}=\tfrac{1}{2}A_0^{-1}\phi_{,N}^2~.
\end{equation}
The effective mass term can be written as
\begin{equation}
    \frac{m^2_{\tilde\rho,{\rm eff}}}{H^2}=\frac{4}{f^2}\hat\xi(3-\epsilon_\phi)\Bigg\{\frac{(1-x)^2-\hat\lambda/\hat\xi}{(1-x)^2-\hat\lambda}-\frac{\epsilon_\phi}{2(3-\epsilon_\phi)}\Bigg\}~,
\end{equation}
where the evolution of $\phi$ is given by its classical EOM. 

If the noise is assumed to be Gaussian,
it suffices to consider quadratic expectation values.
From \eqref{rho_pi_eqs_L}, we obtain their equations of motion:
\begin{align}
\begin{aligned}
    \langle\rho^2\rangle_{,N} &= 2Q+\langle\xi^2_\rho\rangle~,\\
    Q_{,N} &= -\frac{m^2_{\tilde\rho,{\rm eff}}}{H^2}\langle\rho^2\rangle-(3-\epsilon_\phi)Q+\langle{\pi^\rho}^2\rangle+\langle\xi_\rho\xi_{\pi^\rho}\rangle+\langle\xi_{\pi^\rho}\xi_\rho\rangle~,\\
    \langle{\pi^\rho}^2\rangle_{,N} &=-2\frac{m^2_{\tilde\rho,{\rm eff}}}{H^2}Q-2(3-\epsilon_\phi)\langle{\pi^\rho}^2\rangle+\langle\xi^2_{\pi^\rho}\rangle~,\label{FP_eqs}
\end{aligned}
\end{align}
where $Q\equiv\tfrac{1}{2}(\langle\rho\pi^\rho\rangle+\langle\pi^\rho\rho\rangle)$, and $\langle\xi_I\xi_J\rangle$ (with $I,J=\rho,\pi^\rho$) are quadratic (time-dependent) amplitudes of the noise correlators, defined as
\begin{equation}
    \langle\xi_I(N)\xi_J(N')\rangle\equiv\langle\xi_I\xi_J\rangle\delta(N-N')~.
\end{equation}
In terms of the Fourier modes, the noise correlators are given by
\begin{gather}
\begin{gathered}
    \langle\xi^2_\rho\rangle=\frac{|\rho_{k_s}|^2}{2\pi^2}k_s^3|1-\epsilon_\phi|~,~~~\langle\xi^2_{\pi^\rho}\rangle=\frac{|\pi^{\rho}_{k_s}|^2}{2\pi^2}k_s^3|1-\epsilon_\phi|~,\\
    \langle\xi_\rho\xi_{\pi^\rho}\rangle+\langle\xi_{\pi^\rho}\xi_\rho\rangle=\frac{\rho_{k_s}(\pi^{\rho}_{k_s})^*+{\rm c.c.}}{2\pi^2}k_s^3|1-\epsilon_\phi|~.
\end{gathered}
\end{gather}
The noise amplitudes can be derived by using the Fourier mode solution $\rho_k$. In the slow-roll limit, we can approximate it as~\footnote{Numerically, we find that even when slow-roll is temporarily violated near the critical point, the slow-roll expression of $\rho_k$ is still a good approximation for the noise terms in the stochastic equations. Furthermore, quantum diffusion in our models is followed by the second slow-roll stage, which washes out any dependence on the initial conditions, as long as they are not too large.}
\begin{equation}
    \rho_k\simeq \sqrt{A_0}\frac{e^{-ik\tau}}{a\sqrt{2k}}(1+iaH/k)~,~~~\pi^{\rho}_k=\rho_{k,N}~,
\end{equation}
where $\sqrt{A_0}$ appears after the canonical normalization of $\rho$. This leads to the quadratic noise amplitudes,
\begin{gather}
\begin{gathered}
    \langle\xi^2_\rho\rangle\simeq\frac{A_0H^2}{4\pi^2}~,~~~\langle\xi^2_{\pi^\rho}\rangle\simeq \frac{A_0H^2}{4\pi^2}(\epsilon_\phi+s^2)^2~,\\
    \langle\xi_\rho\xi_{\pi^\rho}\rangle+\langle\xi_{\pi^\rho}\xi_\rho\rangle\simeq -\frac{A_0H^2}{2\pi^2}(\epsilon_\phi+s^2)~.\label{noise_amp_SR}
\end{gathered}
\end{gather}
Since $s,\epsilon_\phi\ll 1$ during slow-roll, $\langle\xi^2_\rho\rangle$ becomes the dominant noise term, and the rest can be ignored. In the cases where slow-roll is violated near $N_c^{\rm eff}$, expressions \eqref{noise_amp_SR} can be treated as order-of-magnitude estimates.

We numerically solve the system \eqref{FP_eqs} starting from $N_c^{\rm eff}$, with vanishing initial conditions for $\langle\rho^2\rangle$, $Q$, and $\langle{\pi^\rho}^2\rangle$. We then compute the duration $\Delta N_q$ of the quantum diffusion stage at the end of which, the solution ``classicalizes", viz., the stochastic noise term becomes negligible compared to the field momentum,
\begin{equation}
    \langle{\pi^\rho}^2\rangle=\kappa^2\langle\xi^2_\rho\rangle~,
\end{equation}
for some constant $\kappa\gg 1$ (for numerical solutions we will use $\kappa=100$). After the solution $\sqrt{\langle\rho^2\rangle}$ is found, its value at the time $N_c^{\rm eff}+\Delta N_q$ (as well as its velocity) is taken as the initial condition for the second classical stage of inflation.

\begin{figure}
\centering
  \centering
  \includegraphics[width=.65\linewidth]{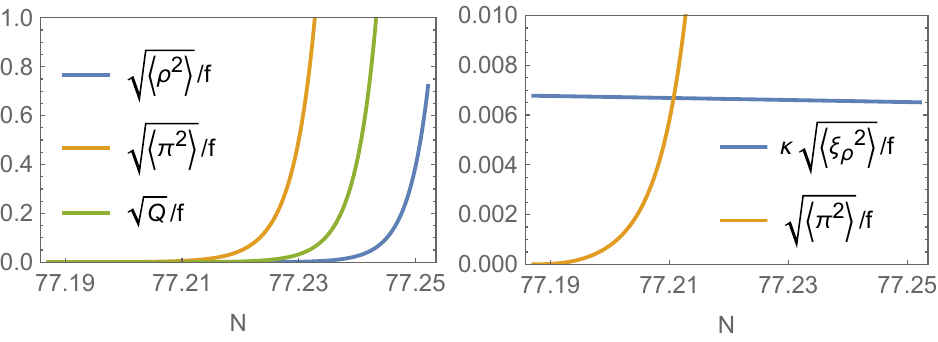}
\captionsetup{width=1\linewidth}
\caption{Solution to stochastic equations \eqref{FP_eqs} (here $\pi$ stands for $\pi^\rho$). The right plot shows how the average momentum becomes much larger than the noise term, by choosing for example $\kappa=100$.}\label{Fig_QD_sols}
\end{figure}

Figure \ref{Fig_QD_sols} showcases a solution to stochastic equations \eqref{FP_eqs} for the parameters $\hat\lambda=1$, $M=8.62\times 10^{-5}$, $\lambda=0.1$ ($f$ is then derived as $f\approx 0.022$). Here, we first run the numerical solution to the background equation \eqref{phi_EOM_full} for $\phi$ (when $\rho=0$) starting from $N=0$ until the effective critical point $N_c^{\rm eff}\approx 77.19$ plus e.g. an additional efold, in order to obtain the $\phi$-dependent coefficients in stochastic equations \eqref{FP_eqs}. The solution to \eqref{FP_eqs} (Fig. \ref{Fig_QD_sols}, left) shows that $\langle{\pi^\rho}^2\rangle$ value starts to grow first, followed by $Q$ and $\langle\rho^2\rangle$. The plot on the right indicates the ``classicalization" time when the momentum average crosses $\kappa\sqrt{\langle\xi^2_\rho\rangle}$, where we set $\kappa=100$. This yields $\Delta N_q\approx 0.024$ in the current example. From this point, we proceed to the second classical stage of inflation.

\subsection{Second stage and numerical background solutions}

Here, we obtain full numerical trajectories for the background system \eqref{phi_EOM_full}--\eqref{V_final}. To summarize the strategy, we divide the inflationary solution into three parts: 1) the first classical solution running from $N=0$ until the effective critical point $N_c^{\rm eff}$; 2) the quantum diffusion stage lasting $\Delta N_q$ efolds, as explained above; 3) the second classical stage taking place between $N_c^{\rm eff}+\Delta N_q$ and until the end of slow-roll, $\epsilon=1$. For the second classical stage, the $\rho$ initial condition is provided by its quantum diffusion stage, while the $\phi$ initial condition can be found from its classical EOM \eqref{phi_EOM_full}, since the quantum diffusion of $\phi$ can be neglected due to its non-vanishing classical velocity (it in fact grows as we approach the critical point).

\begin{figure}
\centering
  \centering
  \includegraphics[width=1\linewidth]{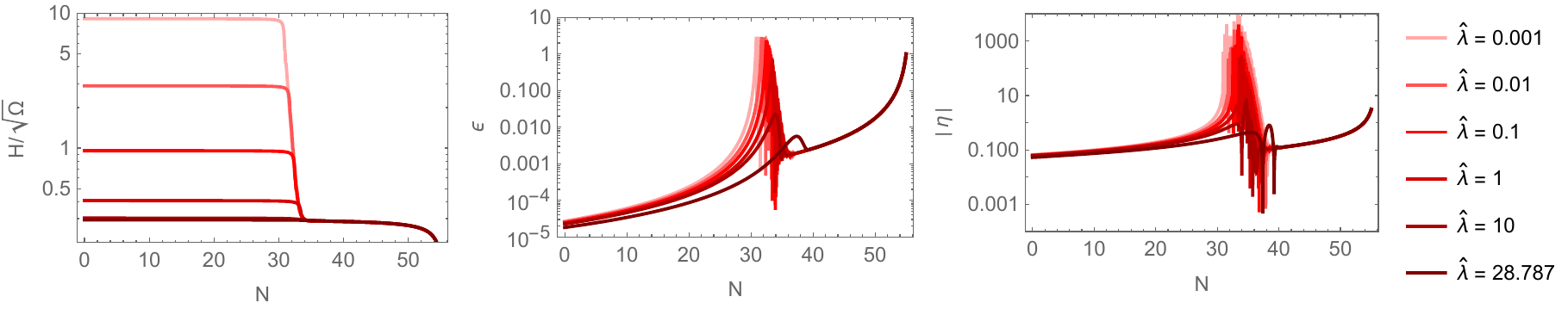}
\captionsetup{width=1\linewidth}
\caption{Hubble function and the slow-roll parameters during the last $55$ efolds of inflation. $H$ is normalized by using $\Omega\equiv\lambda f^4/A_0^2$. Full parameter sets are shown in Table \ref{Tab_pars_xi=30}.}\label{Fig_H_SR}
\end{figure}

We set $\lambda=0.1$ as a reference value, while the inflaton mass parameter $M$ can be fixed from the CMB scalar amplitude \eqref{Amp_CMB}~\footnote{It should be noted that the slow-roll expression \eqref{Amp_CMB} of the amplitude of CMB scalar modes shows the values at the horizon exit, and therefore possible superhorizon changes (due to multi-field effects) can modify the resulting parameter values (particularly, $M$ and $f$ values if we fix $\hat\lambda$ and $\hat\xi$). Superhorizon evolution of perturbations will be studied in the next section, and the parameter values of this section should be understood as preliminary results.}, and we are left with two free parameters $f$ and $\xi$. As mentioned earlier, it is convenient to trade them in favor of the derived parameters
\begin{equation}
    \hat\xi\equiv |\xi|f^2~,~~~\hat\lambda\equiv \frac{\lambda f^4}{3M^2}~.
\end{equation}
In our first set of examples, we will take $\hat\xi=30$, with the expectation that it will lead to $\Delta N_2=20\sim 30$ efolds of PQ inflation, and vary $\hat\lambda$ around unity. The parameter $\hat\lambda$ is bounded from above, $\hat\lambda<\hat\xi$, as required by the positivity of the effective $\rho$ mass \eqref{rho_mass_I} (i.e., restoration of the PQ symmetry) during the first stage of inflation. Figure \ref{Fig_H_SR} shows the Hubble function and the Hubble slow-roll parameters $\epsilon_H$ and $\eta$ for a range of $\hat\lambda_H$, during the last $55$ efolds of inflation ($N=0$ corresponds to the horizon exit of CMB scales). Table \ref{Tab_pars_xi=30} records the rest of the parameters used, as well as the initial conditions from the quantum diffusion stage. We find that as we increase $\hat\lambda$, $\delta N_2$ gradually decreases until we get close to the upper bound $\hat\lambda=\hat\xi=30$, at which it starts to rapidly increase. At $\hat\lambda\approx 28.8$, the second stage lasts more than $53$ efolds. At the same time the duration of the quantum diffusion stage $\Delta N_q$ increases with $\hat\lambda$, and at $\hat\lambda\approx 28.8$ it reaches $\sim 1.5$ efolds. 

From Fig. \ref{Fig_H_SR}, we can see the step-like behavior of the Hubble function between the two slow-roll stages for small $\hat\lambda$, while for large $\hat\lambda$, the Hubble function is smooth and flat until the end of the second stage. By looking at the slow-roll parameters (Fig. \ref{Fig_H_SR}, center and right), we can see spikes and a brief period of oscillations between the two slow-roll phases, where the oscillations are more violent for smaller values of $\hat\lambda$, and slow-roll can be temporarily broken. Nevertheless, the choice $\hat\lambda\approx 28.8$ should be distinguished here because the spike in the slow-roll parameters in this case does not indicate the transition to the second stage as we have defined it, but happens during the second stage (the seamless transition to the second stage happens near $N=0$). 

\begin{figure}
\centering
  \centering
  \includegraphics[width=.9\linewidth]{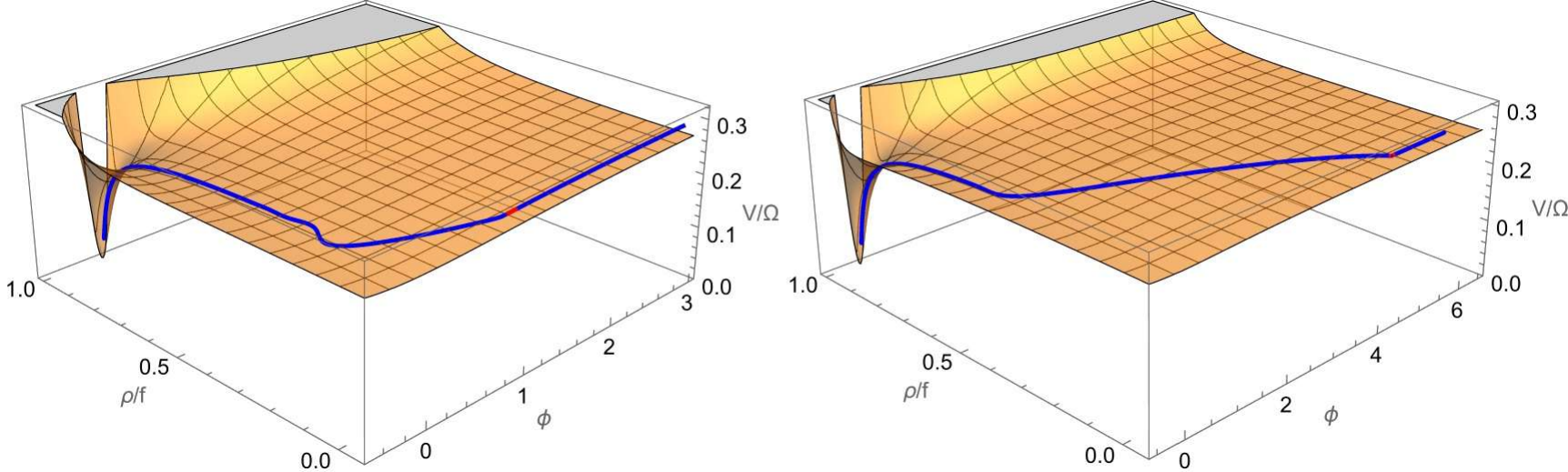}
\captionsetup{width=1\linewidth}
\caption{Inflationary trajectories on the scalar potential for $\hat\lambda=10$ (left) and $\hat\lambda=28.787$ (right). $\Omega\equiv \lambda f^4/A_0^2$. Red portions represent the transient stage.}\label{Fig_3d_compare}
\end{figure}

Let us elaborate on this behavior below by comparing the field trajectories for e.g. $\hat\lambda=10$ with $\hat\lambda=28.787$, as can be seen in Fig. \ref{Fig_3d_compare}, where the classical trajectories are shown in blue, and the short quantum diffusion stage in red. The plot on the left depicts the inflationary trajectory for $\hat\lambda=10$: the start of the second inflationary stage (right after the red portion of the trajectory) is followed by a short transition period where the $\phi-\rho$ trajectory is diagonal. Subsequently, we can see a few oscillations of $\phi$ and finally the $\rho$-driven inflation where $\phi=0$. In contrast, in the extreme case of $\hat\lambda=28.787$ (right-side plot), the transition period with the diagonal part of the trajectory is much longer and lasts from $N\approx 0.1$ until the peak in the slow roll parameters shown in Fig. \ref{Fig_H_SR}, which coincides with the onset of the $\rho$-driven stage. In summary, when $\hat\lambda$ is close to $30$, the second stage can be further divided into the two-field part, and the effectively single-field, $\rho$-driven part. For smaller values of $\hat\lambda$, the two-field part is short. In our examples, the $\rho$-mass around the true (Minkowski) vacuum varies between $m_{\tilde\rho}\sim 10^{-7}$ (for $\hat\lambda=0.001$) and $m_{\tilde\rho}\sim 10^{-6}$ (for $\hat\lambda\approx 28.8$), viz., a few orders below the inflaton mass given by $M$ (the effective $\rho$-mass during the first stage is larger, between $10^{-2}$ and $10^{-5}$, respectively, so $\rho$ is strongly stabilized).

\begin{table}[hbt!]
\centering
\begin{tabular}{l | r r r r r r}
\toprule
$\hat\lambda$ & $0.001$ & $0.01$ & $0.1$ & $1$ & $10$ & $28.787$\\
$M$ & $1.2\times 10^{-4}$ & $1.27\times 10^{-4}$ & $1.19\times 10^{-4}$ & $8.62\times 10^{-5}$ & $3.62\times 10^{-5}$ & $1.97\times 10^{-5}$\\
$f$ & $4.56\times 10^{-3}$ & $8.34\times 10^{-3}$ & $1.44\times 10^{-2}$ & $2.17\times 10^{-2}$ & $2.5\times 10^{-2}$ & $2.41\times 10^{-2}$\\
$\hat\rho_i$ & $3.97\times 10^{-5}$ & $3.28\times 10^{-5}$ & $2.77\times 10^{-5}$ & $4.05\times 10^{-5}$ & $1.44\times 10^{-4}$ & $1.42\times 10^{-3}$\\
$\partial_N\hat\rho_i$ & $2.76\times 10^{-2}$ & $1.6\times 10^{-2}$ & $9.32\times 10^{-3}$ & $6.68\times 10^{-3}$ & $6.67\times 10^{-3}$ & $6.38\times 10^{-3}$\\
$\Delta N_q$ & $0.004$ & $0.006$ & $0.01$ & $0.024$ & $0.107$ & $1.496$\\
$\Delta N_2$ & $24.169$ & $23.316$ & $22.572$ & $21.6$ & $21.427$ & $53.408$\\
\bottomrule
\end{tabular}
\captionsetup{width=1\linewidth}
\caption{Parameters sets (for $\hat\xi=30$) and the $\rho$ initial conditions ($\hat\rho\equiv \rho/f$) obtained from the stochastic equations. The last two rows show the duration of the quantum diffusion stage $\Delta N_q$ and the second classical stage of inflation $\Delta N_2$.}
\label{Tab_pars_xi=30}
\end{table}

\begin{figure}
\centering
  \centering
  \includegraphics[width=1\linewidth]{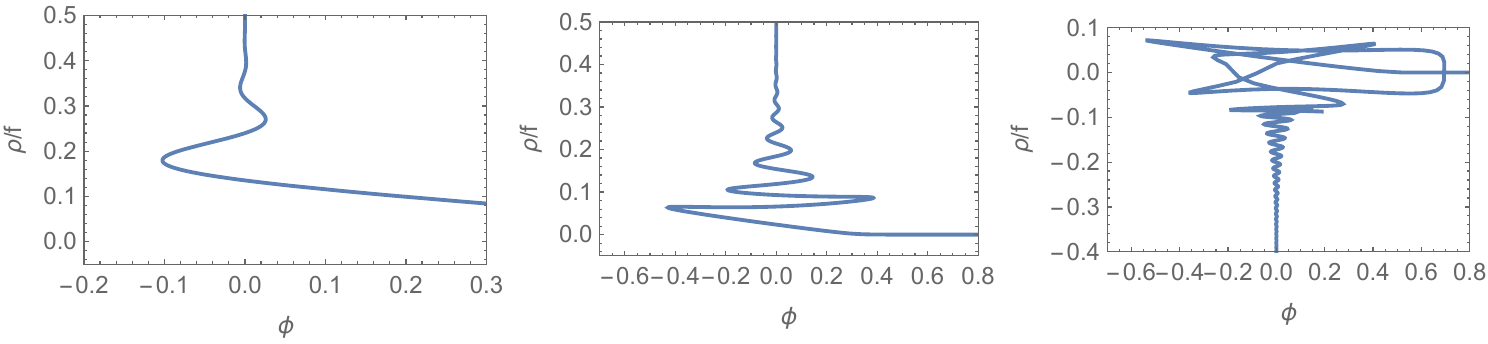}
\captionsetup{width=1\linewidth}
\caption{Field-space trajectories for $\hat\lambda=10,1,0.1$ (from left to right).}\label{Fig_phi_rho_sols}
\end{figure}

\begin{figure}
\centering
  \centering
  \includegraphics[width=.4\linewidth]{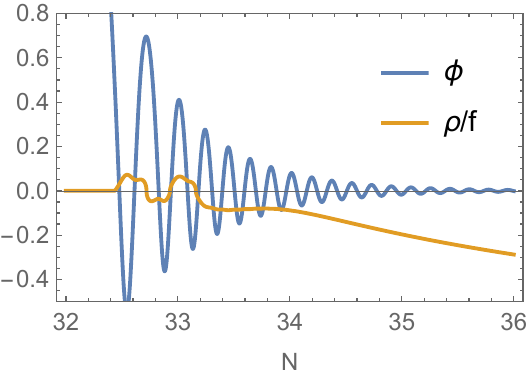}
\captionsetup{width=1\linewidth}
\caption{Oscillations of $\phi$ and $\rho$ at the start of the second stage, $\hat\lambda=0.1$.}\label{Fig_phi_rho_osc}
\end{figure}

The field-space trajectories for $\hat\lambda=10,1,0.1$ can reveal the origin of the oscillations in the slow-roll parameters. Figure \ref{Fig_phi_rho_sols} shows the trajectories as we decrease $\hat\lambda$, around the transition to the second stage. The plot on the left shows few oscillations of $\phi$, while $\rho$ gradually grows until it reaches its VEV $\langle\rho\rangle=f$. For $\hat\lambda=1$ (center plot), $\phi$ oscillates faster, but $\rho$ still increases smoothly. For $\hat\lambda=0.1$ (right plot) the oscillations of $\phi$ further increase, but now $\rho$ also starts to oscillate around the origin (see Fig. \ref{Fig_phi_rho_osc}), which results in a ``chaotic" behavior of the trajectory where $\rho$ goes from positive to negative values before the trajectory stabilizes around $\phi=0$, and $\rho$ finally reaches its (negative) VEV. Note that the negative values of $\rho$ should be unphysical by the definition of $\rho$ as the radial direction of the complex PQ field $S$. Nevertheless, these negative excursions of $\rho$ can be understood as the angular change of $S$, i.e., the change in the axion value after the PQSB. This can be seen by parametrizing $S$ as real and imaginary parts, $S=S_1+iS_2$. Since the EOM are $U(1)_{\rm PQ}$-symmetric, this yields identical equations for the scalars $S_1$ and $S_2$, and the negative values in the trajectory of $\rho$ can be understood as negative values of both $S_1$ and $S_2$ (we confirm this by numerically solving the system with scalars $S_1$ and $S_2$ for the parameter choice $\hat\lambda=0.1$ in Fig. \ref{Fig_phi_rho_sols}).

\begin{figure}
\centering
  \centering
  \includegraphics[width=.55\linewidth]{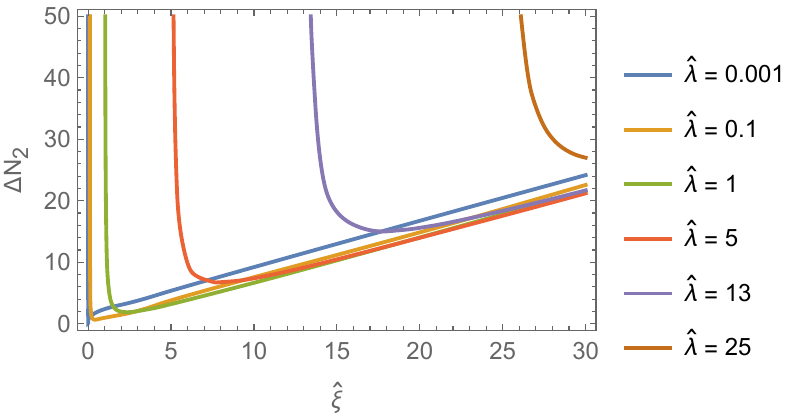}
\captionsetup{width=1\linewidth}
\caption{$\Delta N_2$ as a function of $\hat\xi$ for a range of $\hat\lambda$.}\label{Fig_DN_2_xi}
\end{figure}

To see how $\hat\xi$ controls the duration $\Delta N_2$ of the second stage, we plot $\Delta N_2$ as a function of $\hat\xi$ for a given $\hat\lambda$ in Fig. \ref{Fig_DN_2_xi}. For a given $\hat\lambda$, $\hat\xi$ is bounded as $\hat\xi>\hat\lambda$ by the positivity of the $\rho$ mass-squared \eqref{rho_mass_I} during the first stage, and $f$ is varied between $\sim 10^{-3}$ and $\sim 10^{-2}$ (increases with $\hat\lambda$) in the given examples. Almost linear dependence can be seen when $\hat\xi\gg\hat\lambda$, as expected from our analysis of the single field (non-minimally coupled) PQ inflation in Section \ref{sec_NM_PQ} (see Eq. \eqref{N_estimate_NM_PQ} for the corresponding number of efolds). Nonetheless, when $\hat\xi$ approaches $\hat\lambda$ from above, a multi-field effect is seen: $\Delta N_2$ undergoes steep growth, which is steeper for smaller values of $\hat\lambda$. As a consequence, $\Delta N_2$ has a minimum for a given $\hat\lambda$, and this minimum increases with $\hat\lambda$. We have seen this sudden growth of $\Delta N_2$ in the case of $\hat\lambda\approx 28.8$ and $\hat\xi=30$, and the multi-field trajectory responsible for the large $\Delta N_2$ can be seen in Fig. \ref{Fig_3d_compare} (right plot).

\section{Scalar perturbations}\label{Sec_perturbations}

Let us now turn to scalar perturbations and their superhorizon evolution due to multi-field dynamics. We will work with gauge-invariant perturbations
\begin{equation}
    \delta\Phi_g^A=\delta\Phi^A+\dot\Phi^A\psi/H~,
\end{equation}
where $\delta\Phi^A$ is the perturbation of the $\Phi^A$ field, and $\psi$ is the perturbation of the spatial part of the (spacetime) metric tensor. In the covariant formalism (see e.g. \cite{Kaiser:2012ak,Gundhi:2020kzm}), we can write the EOM for $\delta\Phi^A$ as (from now we will drop subscript ``$g$")
\begin{equation}
    D_t^2\delta\Phi^A+3HD_t\delta\Phi^A+\Big(\frac{k^2}{a^2}\delta^A_B+{M^A}_B\Big)\delta\Phi^B=0~,
\end{equation}
where the field-space covariant derivative along the time direction $D_t$ of a vector $V^A$ is
\begin{equation}
    D_tV^A=\dot V^A+\dot\Phi^B\Gamma^A_{BC}V^C~,
\end{equation}
and the effective mass matrix ${M^A}_B$ is given by
\begin{equation}
    {M^A}_B=\nabla^A\nabla_BV+{R^A}_{CBD}\dot\Phi^C\dot\Phi^D-a^{-3}D_t(a^3\dot\Phi^A\dot\Phi_B/H)~.
\end{equation}
From the field basis we can switch to the adiabatic-isocurvature basis, which for the two-field case consists of an adiabatic mode $Q_\sigma$ and a single isocurvature mode $Q_s$:
\begin{equation}
    Q_\sigma=G_{AB}\,e_\sigma^A\delta\Phi^B~,~~~Q_s=G_{AB}\,e_s^A\delta\Phi^B~,
\end{equation}
where the basis unit vectors can be written as
\begin{equation}
    e^A_\sigma=\frac{\dot\Phi^A}{\sqrt{G_{BC}\dot\Phi^B\dot\Phi^C}}~,~~~e^A_s=\frac{D_te_\sigma^A}{\sqrt{G_{BC}D_te_\sigma^B D_te_\sigma^C}}~.
\end{equation}
In terms of $Q_\sigma$ and $Q_s$, the power spectra of curvature and isocurvature perturbations are
\begin{equation}
    P_\zeta=\frac{k^3}{2\pi^2}\,\frac{H^2|Q_\sigma|^2}{G_{AB}\dot\Phi^A\dot\Phi^B}~,~~~P_S=\frac{k^3}{2\pi^2}\,\frac{H^2|Q_s|^2}{G_{AB}\dot\Phi^A\dot\Phi^B}~.
\end{equation}

\begin{figure}
\centering
  \centering
  \includegraphics[width=.85\linewidth]{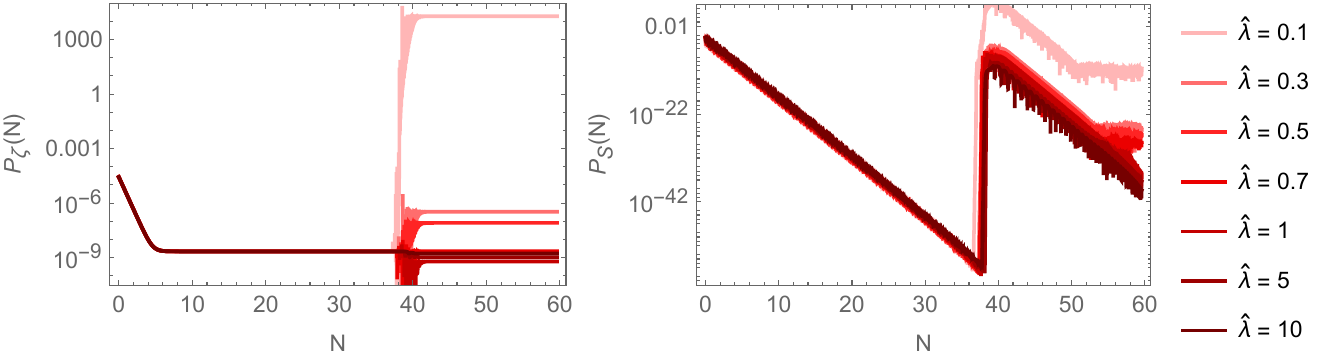}
\captionsetup{width=1\linewidth}
\caption{Superhorizon evolution of the power spectra $P_\zeta$ and $P_S$ for the CMB mode $k_*$, normalized ($P_\zeta\approx 2.1\times 10^{-9}$) at the moment of horizon exit.}\label{Fig_Pk_exit}
\end{figure}

\begin{figure}
\centering
  \centering
  \includegraphics[width=1\linewidth]{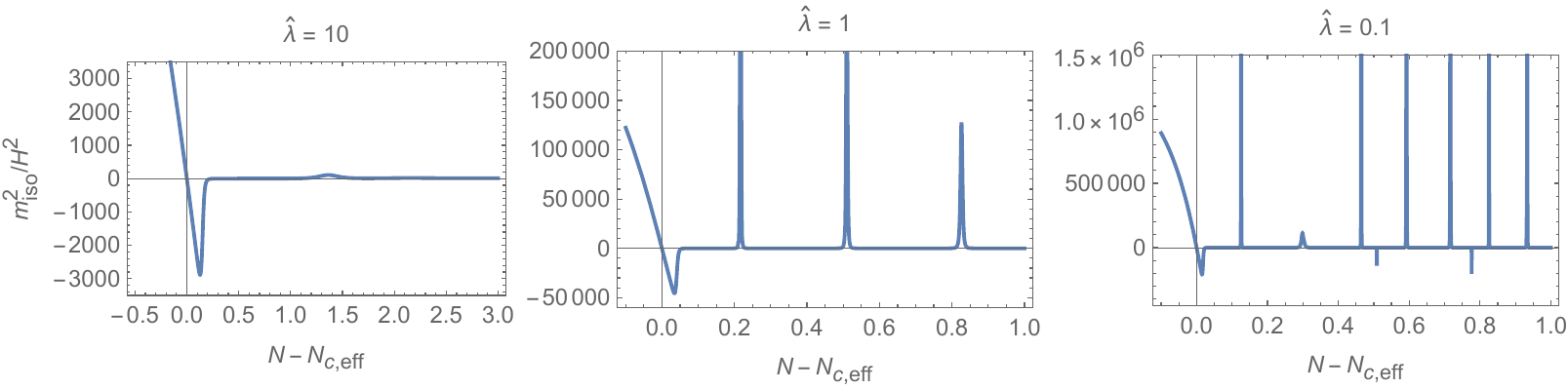}
\captionsetup{width=1\linewidth}
\caption{Effective isocurvature mass for $\hat\lambda=10,1,0.1$. The values of $M$ are shown in Table \ref{Tab_pars_xi=30}.}\label{Fig_m2iso}
\end{figure}

For the case $\hat\xi=30$ with different values of $\hat\lambda$, Fig. \ref{Fig_Pk_exit} plots the superhorizon changes of the power spectrum by first considering the CMB scale $k_*$ (we evolve the perturbation from around $5$ efolds before the horizon exit until the end of inflation). For $\hat\lambda\gtrsim 1$, our testing shows relatively stable behaviour of the $k_*$ mode, but for small $\hat\lambda$, significant amplification can be seen after rapid oscillations around the beginning of the second stage (numerical computation time for small $\hat\lambda$ also becomes increasingly longer, and prone to stiffness errors). This amplification is likely due to the spikes in the isocurvature effective mass caused by the oscillatory evolution of the fields:  \cite{Kaiser:2012ak,Braglia:2020eai,Gundhi:2020kzm,Braglia:2020taf}
\begin{equation}
    m_{\rm iso}^2=G_{AC}e_s^Ce_s^B{m^A}_B+3G_{AB}D_te_{\sigma}^AD_te_{\sigma}^B~,
\end{equation}
where
\begin{equation}
    {m^A}_B=\nabla^A\nabla_BV+{R^A}_{CBD}\dot\Phi^C\dot\Phi^D~.
\end{equation}
For $\hat\lambda=10,1,0.1$ (and $\hat\xi=30$), we plot $m_{\rm iso}^2/H^2$ in Fig. \ref{Fig_m2iso}, where the narrow upward (positive) spikes are already evident for $\hat\lambda=1$. The spikes become increasingly higher and narrower, while for $\hat\lambda=0.1$, the isocurvature mass-squared develops downward (negative) spikes as well.

Regarding the normalization of the CMB modes for small $\hat\lambda$ (Fig. \ref{Fig_Pk_exit}) at $\hat\xi=30$, decreasing the parameter $M$ (and hence $f$) in order to decrease the amplitude at the end of inflation (to match $P_\zeta(k_*)\approx 2.1\times 10^{-9}$) is not straightforward because of the non-trivial superhorizon evolution, unlike what one would expect from usual slow-roll models. For example, monotonically decreasing $M$ can result in sudden changes (both positive and negative) in the final amplitude $P_\zeta(k_*,N_{\rm end})$, due to the highly oscillatory transient regime between the two slow-roll stages. Such high oscillations require more sophisticated numerical methods to deal with stiffness and lengthy computation times. On the other hand, because of the sensitivity of the final amplitude on the oscillatory phase, possible particle production (introducing additional friction term) or change in the shape of the scalar potential around the transient stage may significantly modify the superhorizon evolution of the perturbations. For $\hat\lambda\gtrsim\co(1)$ and especially for larger values, the results and inflationary trajectories in general are less model dependent (in terms of the inflaton sector) due to the definition of $\hat\lambda\equiv\lambda f^4/(3M^2)$ as the ratio between the PQ potential at $\rho=0$ and the inflaton potential at the plateau region (large $\hat\lambda$ means that specific shape of the inflaton potential becomes less important).

\begin{figure}
\centering
  \centering
  \includegraphics[width=.85\linewidth]{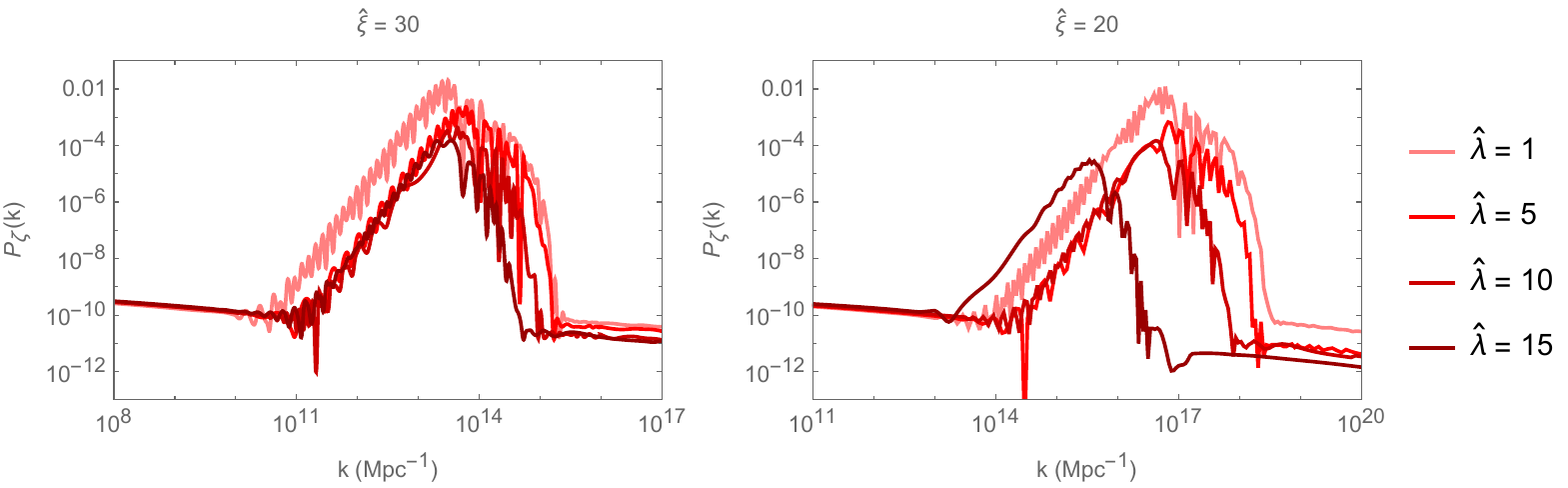}
\captionsetup{width=1\linewidth}
\caption{Scalar power spectrum for a range of $\hat\lambda$ and $\hat\xi=30$ (left) and $\hat\xi=20$ (right).}\label{Fig_Pofk}
\end{figure}

Finally, we address the power spectrum and the amplification of perturbations at small scales. Figure \ref{Fig_Pofk} plots $P_\zeta(k)$ for $\hat\xi=30$ and $\hat\xi=20$, and several values of $\hat\lambda$~\footnote{The values of $M$ have been fixed for these plots after taking into account superhorizon evolution of the CMB mode: $M=\{1.21\times 10^{-4},7.4\times 10^{-5},3.62\times 10^{-5},3.1\times 10^{-5}\}$ for $\hat\lambda=\{1,5,10,15\}$, respectively.}. The peaks in the power spectrum are located approximately at the critical values of $k$, viz., at the scales exiting the horizon at the critical time $t_c^{\rm eff}$ when the effective mass of $\rho$ vanishes (and the amplification of each nearby $k$-mode happens right after the second classical stage begins). For $\hat\lambda<1$, we find even larger peaks in the power spectrum, and already at $\hat\lambda=0.5$, the peak reaches around $P_\zeta\sim 0.1$ (we do not plot the whole power spectrum for these cases because of long numerical computation times). By computing the resulting PBH abundance (using the Press--Schechter formalism \cite{Press:1973iz,Inomata:2017okj}), we find overproduction of PBH for $\hat\lambda\sim 0.5$ or smaller, and exclude these values.

\begin{figure}
\centering
  \centering
  \includegraphics[width=.6\linewidth]{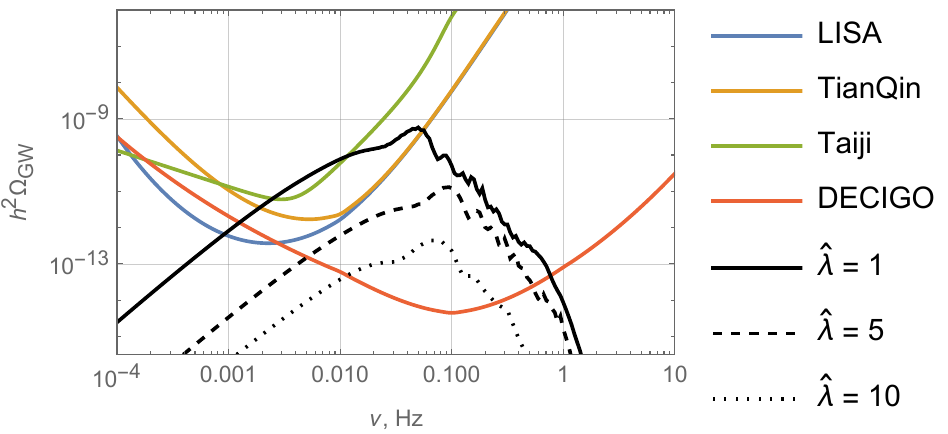}
\captionsetup{width=1\linewidth}
\caption{Scalar-induced gravitational wave density spectrum for the scalar power spectra of Fig. \ref{Fig_Pofk} (with $\hat\xi=30$).}\label{Fig_GW}
\end{figure}

For the power spectra shown in Fig. \ref{Fig_Pofk}, while the PBH abundance is found to be negligible, scalar-induced gravitational waves can be significant. Scalar-induced GW, generated at the second order ($P_h\sim P^2_\zeta$) can be computed for a given scalar power spectrum by using the method described in Refs. \cite{Espinosa:2018eve,Kohri:2018awv}.Figure \ref{Fig_GW} depicts the scalar-induced GW density $\Omega_{\rm GW}$ ($h$ is the reduced Hubble parameter $h=0.67$) as a function of frequency $\nu$. It shows the stochastic GW background reaching the sensitivities of the upcoming spaceborne interferometers, such as LISA \cite{LISA:2017pwj}, TianQin \cite{TianQin:2015yph}, Taiji \cite{Ruan:2018tsw}, and DECIGO \cite{Kudoh:2005as}. The GW densities are shown for $\hat\xi=30$, corresponding to $\Delta N_2\sim 22$ efolds. For smaller values of $\hat\xi$ (shorter second stage), the curves move to higher frequencies (for $\hat\xi=20$ they are outside of the shown interferometer sensitivities). Likewise, larger $\hat\xi$ will prolong the second stage and decrease the peak frequency. We also confirm that first-order primordial GW are not amplified and can be neglected in comparison to scalar-induced GW at the amplified scales.

\section{Discussion}\label{Sec_discussion}

In this work, we considered a cosmological scenario where PQ symmetry breaking occurs in the middle of inflation due to the non-minimal coupling to gravity. The early stage of inflation is driven by an inflaton field $\phi$, while the non-minimal coupling of the PQ field $\rho$ ensures its stability at the symmetric point. As inflation goes on, the effective PQ mass turns tachyonic, and spontaneous symmetry breaking is triggered by the quantum diffusion of $\rho$, which we analyze by using stochastic formalism. After a short period of time (typically $\Delta N\ll 1$), the momentum/velocity of $\rho$ becomes much larger than the stochastic noise, and classical equations of motion can be used again to describe the second slow-roll stage, with initial conditions provided by the stochastic equations. Since slow-roll resumes during the second stage, the inflationary trajectories are largely insensitive to these initial conditions. The main ingredient that makes inflation in the $\rho$-direction (i.e., the second stage) possible is the negative non-minimal coupling, $\xi<0$, with large enough magnitude. More specifically, the duration of the second stage $\Delta N_2$ grows almost linearly with $\hat\xi\equiv|\xi|f^2$ as shown in Fig. \ref{Fig_DN_2_xi}, aside from extreme cases where $\hat\xi$ is close to $\hat\lambda\equiv \lambda f^4/(3M^2)$. In such cases, the $\Delta N_2$ grows extremely fast as $\hat\xi$ approaches $\hat\lambda$ from above (restoration of the PQ symmetry during the first stage requires $\hat\xi>\hat\lambda$). Another important consequence of our scenario is the existence of the minimal duration $\Delta N_2$ of the second stage: for each given $\hat\lambda$, $\Delta N_2$ as a function of $\hat\xi$ has a non-zero minimum proportional to the value of $\hat\lambda$. In other words, by using the non-minimal coupling to restore PQ symmetry during the $\phi$-driven inflation, we automatically get non-zero $\Delta N_2$, viz meso-inflationary PQSB. For small $\hat\lambda$, the minimal $\Delta N_2$ is small, but for $\hat\lambda\gtrsim 1$, it grows proportionally.

The parameter $\hat\lambda$ measures the ratio of the PQ potential to the inflaton potential at the slow-roll plateau, and can be used to outline three distinct parameter regimes: case (a) where $\hat\lambda\sim\co(1)$ and both potentials are close in magnitude; case (b) where $\hat\lambda\ll 1$ and the inflaton potential dominates; case (c) where $\hat\lambda\gg 1$ and the PQ potential dominates. In cases (a) and (c), the axion decay constant $f$ is constrained from below by the observed amplitude of CMB scalar fluctuations, and in combination with experimental constraints, we have $10^{-3}\lesssim f\lesssim 10^{-1}$ in Planck units, i.e., a GUT-scale decay constant. In case (b), the decay constant can in principle be arbitrarily small. In transition between the two stages of inflation, slow-roll can be badly violated in case (b) due to large oscillations of both scalar fields but mildly in cases (a) and (c) (e.g., $\epsilon_H$ can be small while $\eta_H\geq 1$) or absent.

In all three cases,
the brief transient stage of quantum
diffusion (between stages I and II)
plays a vital role
as it sets the initial conditions
for the second stage of our model
(see e.g. Table \ref{Tab_pars_xi=30}).
During the transient stage,
the classical force from the potential
is small enough that the quantum backreaction of the short-wavelength quantum modes becomes
important for background evolution.
Physically, quantum uncertainty will select out a specific set of $\rho$ and $\partial_N \rho$ values to break the symmetry, via wavefunction collapse.
The entire evolution between
stage I and stage
II concerns a classical-to-quantum and
then a quantum-to-classical process.
A first-principle description of such
evolution is challenging as the physics behind a wavefunction collapse remain elusive.
Instead, we have adopted a strategy
where we apply the stochastic formalism
and approximate
the collapsed field value with its statistical variance, evaluated when the stochastic
noise is much smaller than the classical
force.
For the present work,
this approximation suffices.
But for more generic applications,
say, in the case where non-Gaussianities
are relevant,
one may wish to adopt a more generic
quantum backreaction framework,
such as in \cite{bojowald2020canonical,ding2023effective}.

By studying scalar perturbations in this model, we find that the non-trivial field trajectories at the start of the second stage can lead to superhorizon changes of the power spectrum, including the CMB scales. In particular, the modes that exit the horizon around the effective critical point (when the $\rho$-mass vanishes) is amplified the most, which can be interesting from the perspective of PBHs and GW. We find that for small $\hat\lambda$ (which translates into sub-GUT-scale decay constant), the superhorizon growth of the perturbations can become uncontrollably large, which can be used to exclude this parameter region by PBH overproduction. At the same time, the superhorizon evolution in this case becomes more sensitive to the field oscillation patterns (at the start of the second stage), which depend on the shape of the scalar potential as well as friction terms caused by possible particle production, which can reduce the number of oscillations (we also find that for small $\hat\lambda$, the Mathematica computation times can become extremely lengthy, and other numerical problems can arise). This sensitivity deserves more research, including more sophisticated numerical methods. In contrast, for $\hat\lambda\gtrsim 1$, the field oscillations are mild, while the perturbations are relatively under control and less sensitive to the shape of the scalar potential (namely the inflaton potential). In this case, we find the amplification of the power spectrum at small scales up to $P_\zeta\sim 10^{-2}$ (for $\hat\xi\sim\co(10)$), yet with no significant PBH production. Nonetheless, these power spectra (Fig. \ref{Fig_Pofk}) can generate large second-order GW, as shown in Fig. \ref{Fig_GW}, potentially reaching the sensitivities of the upcoming spaceborne GW experiments.

Let us also comment on the values of the non-minimal coupling $\xi$. The duration of the second stage is roughly set by the value of $\hat\xi$ (aside from the extreme case $\hat\xi\approx\hat\lambda$). Hence, for meso-inflationary PQSB, where $\Delta N_2$ is between $\co(1)$ and $\co(10)$, we need $\hat\xi$ also in that range. Then, the magnitude of the original non-minimal coupling $\xi$ is $|\xi|=c/f^2$, where $c$ sits between $\co(1)$ and $\co(10)$. As was shown, when $\hat\lambda\gtrsim 1$, the CMB normalization leads to the GUT-scale decay constant $f$ of around $\co(10^{-3})\sim \co(10^{-1})$, in which case $|\xi|=\co(10^2)\sim\co(10^7)$. For $\hat\lambda<1$, when $f$ can be much smaller, the magnitude of $\xi$ becomes even larger and may be problematic in view of effective field theory, see, e.g. \cite{Bezrukov:2007ep,Barvinsky:2008ia,Burgess:2009ea,Hertzberg:2010dc}. This is an additional argument favoring large $\hat\lambda$ models, which bear GUT-scale decay constant. On the other hand, the non-minimal coupling of the PQ field may help solve the axion quality problem as discussed in \cite{Hamaguchi:2021mmt}, for the values $|\xi|\gtrsim 2\times 10^3$, which are compatible with our scenario of meso-inflationary PQSB, even for large $\hat\lambda$.

Our models can be considered as hybrid inflation with long waterfall regime. Nevertheless, in contrast to the conventional hybrid inflation with long waterfall \cite{clesse2010hybrid,Clesse:2015wea,Kawasaki:2015ppx,kodama2011on,Braglia:2022phb,tada2023primordial,tada2023stochastic}, we focused on large field inflation (for the first stage), while the long waterfall regime in the PQ direction is realized by the flattening effect of the negative non-minimal coupling of the PQ field (without the need for super-Planckian axion decay constant), which at the same time restores $U(1)_{\rm PQ}$ during the first stage. One can also see some similarities (including the methods used) with the model of \cite{Gundhi:2020kzm}, where the authors studied a scalar field coupled to $R+R^2$-gravity, resulting in effectively a two-field inflation with a transient quantum diffusion phase.

Even though we used the Starobinsky model for the inflaton sector as an example, one can easily generalize it to any single-field inflationary model of interest, while the PQ potential can be considered as a general symmetry-breaking potential, not necessarily relevant for the strong CP problem of QCD.

\section*{Acknowledgements}

DD is supported by China Postdoctoral Science Foundation (Certificate No.2023M730704). YW is supported by the General Program of Science and Technology of Shanghai No. \linebreak 21ZR1406700, and Shanghai Municipal Science and Technology Major Project (Grant No. 2019SHZDZX01). YW is grateful for the hospitality of the Perimeter Institute during his visit, where the main part of this work is done. This research was supported in part by Perimeter Institute for Theoretical Physics. Research at Perimeter Institute is supported by the Government of Canada through the Department of Innovation, Science and Economic Development and by the Province of Ontario through the Ministry of Research, Innovation and Science.

\appendix

\section{Canonical stochastic formalism}\label{App_A}
Here we formulate the canonical (or phase-space)
version standard stochastic formalism.
The reason to repeat the exercise is mainly two-fold.

First, in the context of multi-field inflation
with a brief quantum diffusion phase in between the prolonged classical evolution,
one often uses the Lagrangian and slow-roll version of the Langevin equation.
In this case, only $\rho'(\tau)$ appears in the equations of motion
(where $\tau$ is some yet unspecified parameterization of time).
Schematically, when $\tau$ is chosen to be efold time, it may look something like
\begin{equation}
	\rho'(\tau)\simeq \frac{V'(\rho)}{3H^2}+\kappa \xi\,.
	\label{}
\end{equation}
Here, $\xi$ is a stochastic variable
following a normalized Gaussian statistics
and $\kappa$ characterizes its (typically constant) amplitude.

In this version, the field acceleration and its associated noise
are often assumed to play a weaker role
so as to be disregarded in the equations of motion.
This might not be applicable, say, in cases where there is a large deceleration
of $\rho'(\tau)$ caused by Hubble friction, like in USR.
The canonical momentum term and its noise cannot be
neglected \textit{a priori}.

Second, a rigorous formulation of the stochastic formalism should start from canonical quantization of phase-space variables and then use the Wigner formalism
to obtain an approximate phase-space distribution 
\cite{vennin2020stochastic,martin2016quantum}
to represent stochastic uncertainty.
While one could skip all this and heuristically add noise terms (with presumed statistics) to the equations
of motion,
one would not be able to know when or how to generalize these equations when they fail
to approximate the true quantum evolution,
especially when non-Gaussianites
are not negligible.
(We will not consider non-Gaussian aspects
in this work. But the concern stands in general.)

\subsection{Classical dynamics}
We will now recall key elements of the (canonical version)
of the stochastic formalism. For our purpose, we start with the classical dynamics of a two-field model, with fields $\rho$ and $\phi$, as in the main text. But we will assume quantum diffusion is only non-negligible in the $\rho$-direction of the field space.

As in the main text,
one considers the action (in Einstein frame and $M_P=\sqrt{1/(8\pi G_N)}=1$)
\begin{equation}
	S[\rho,\phi,g_{\mu\nu}]
	=\int d\tau d^3x\sqrt{-g}
	\Big[\frac{R}{2}
	-\frac{1}{2}G_{\rho\rho}\partial\rho\partial\rho
	-\frac{1}{2}G_{\phi\phi}\partial\phi\partial\phi
	-V_1(\phi)-V_2(\rho)
	\Big]\,.
	\label{}
\end{equation}

The metric is assumed to be in an FLRW form with unspecified time parameterization
$\tau$
\begin{equation}
	ds^2=g_{\mu\nu}dx^{\mu}dx^{\nu}
	=-N^2(\tau)d\tau^2+a^2(\tau)\delta_{ij}dx^idx^j\,.
	\label{}
\end{equation}
The lapse function $N(\tau)$ is commonly chosen to be
$N(\tau)=1,a(\tau)$, and $H^{-1}$ for cosmic time, conformal time,
and efold number respectively.
(In the appendix, the function $N(\tau)$ will be reserved for the lapse function
and we will never commit to a specific time parameterization for $\tau$,
so that the efold number --- also denoted as $N$ in the main text ---
will never appear here.)

The Ricci scalar and the square root of the metric determinant give
\begin{equation}
	R=6\Big(\frac{a''}{N(\tau)^2a}
	+\frac{(a')^2}{N(\tau)^2 a^2}-\frac{a'N'(\tau)}{aN(\tau)^3}\Big)\,,\;
	\sqrt{-g}=N(\tau) a^3\,,
	\label{}
\end{equation}
where a prime denotes a derivative with respect to the argument of the function.
The purely gravitational part of the action can thus be simplified with
integration-by-parts
\begin{equation}
	S_{\RM{grav}}\equiv \frac{1}{2}\int d\tau d^3x
	\sqrt{-g} R=-3\int d\tau d^3x\frac{a(a')^2}{N(\tau)}\,.
	\label{}
\end{equation}

The canonical conjugates are thus
\begin{equation}
	\Pi_a=-\frac{6a}{N(\tau)}a'\,,\; 
	\Pi_{\rho}=\frac{a^3 G_{\rho\rho}}{N(\tau)}\rho'\,,\; 
	\Pi_{\phi}=\frac{a^3 G_{\phi\phi}}{N(\tau)}\phi'\,,\;
	\Pi_N=0\,.
	\label{}
\end{equation}

To obtain the Hamiltonian, first
recall that the lack of $N'(\tau)$ in the Lagrangian gives rise to the
primary constraint $\Pi_N=0$.
This requires us to introduce a Lagrangian multiplier $\lambda$,
giving us the Hamiltonian
\begin{align*}
	H &=\int d^3x(
	\mathcal{H}_g+\mathcal{H}_m)\\
	\mathcal{H}_g
	 &=-\frac{N(\tau)}{12 a}\Pi_a^2+\lambda \Pi_N\\
	\mathcal{H}_m
	 &=\frac{1}{2}
	\Big(
	\frac{N(\tau)}{a^3 G_{\rho\rho}}\Pi_{\rho}^2
	+N(\tau)a G_{\rho\rho}\partial_i\rho\partial_j\rho\delta^{ij}\\
	&+
	\underbrace{\frac{N(\tau)}{a^3 G_{\phi\phi}}\Pi_{\phi}^2
	+N(\tau)a G_{\phi\phi}\partial_i\phi\partial_j\phi\delta^{ij}
	+N(\tau)a^3(V_1+V_2)}_{N(\tau)U}\Big)\,,
	\label{}
\end{align*}
where we have defined the function $U\equiv U(\rho,\phi,\Pi_{\phi})$
to separate the $\rho$-kinetic terms from the rest of the matter Hamiltonian
(density).

Finally, the classical dynamics are given by the Hamilton's equations.
The gravity part reads
\begin{align*}
	\Pi_a &=-\frac{6a}{N(\tau)}a'\,,\; 
	\lambda=N'(\tau)\\
	\Pi_N' &=\frac{1}{12a}\Pi_a^2-\frac{\partial \mathcal{H}_m}{\partial N(\tau)}
	=0 \text{ (from constraint }
	\Pi_N=0)\\
	\Pi_{a}' &=
	-\frac{N(\tau)}{12 a^2}\Pi_a^2-\frac{\partial\mathcal{H}_m}{\partial a}\,.
\end{align*}
The $N'(\tau)$-equation reflects the time-reparameterization freedom
due to the choice of the function $\lambda(\tau)$,
and the second line is the familiar ``$H^2$-equation'' of the Friedmann equations.

The matter part reads
\begin{align}	
	\Pi_{\rho} &=\frac{a^3 G_{\rho\rho}}{N(\tau)}\rho'\,,\; 
	\Pi_{\phi}=\frac{a^3 G_{\phi\phi}}{N(\tau)}\phi'\nonumber\\
	\Pi'_{\rho} &=
	\frac{N(\tau)}{2a^3G^2_{\rho\rho}}\partial_{\rho}G_{\rho\rho}\Pi_{\rho}^2
	+N(\tau)(aG_{\rho\rho}\partial_i\partial_j\rho\delta^{ij}
	-\frac{a}{2}\partial_{\rho}G_{\rho\rho}\partial_i\rho\partial_j\rho
	\delta^{ij}-
	\partial_{\rho}U)\nonumber\\
	\Pi_{\phi}' &=N(\tau)a G_{\phi\phi}\partial_{i}\partial_{j}\phi\delta^{ij}
	-N(\tau)a^3 \partial_{\phi}(V_1+V_2)\,.
	\label{matter-eom-canonical}
\end{align}

\subsection{Backreaction: stochastic corrections}
The fields $\rho$ and $\phi$ of the Hamiltonian above can be canonically quantized. Nonetheless, since the field couples to the geometry via the Friedmann equations, quantizing the full system concerns quantum gravity and is beyond our present reach. Instead, we split the system into large and small scales separated
by a time-dependent cut-off scale 
$r_{\sigma}=1/k_{\sigma}=(\sigma aH)^{-1}$
with constant $\sigma\ll 1$. We quantize the small-scale modes with wavenumbers above $k_{\sigma}$ and view their backreaction on the large-scale modes effectively
as noise.

More concretely, the splitting is achieved with the help of a window function $W(x)$
such that $W(x)\sim 1$ for $x\gg 1$ and $W(x)\sim 0$ when $x\ll 1$:
\begin{align}
	\rho &=\bar{\rho}(\tau,\mathbf{x})+\rho_s(\tau,\mathbf{x})\,,\;
	\Pi_{\rho}=\bar{\Pi}(\tau,\mathbf{x})+\Pi_s(\tau,\mathbf{x})\nonumber\\
	\rho_s(\tau,\mathbf{x})
	 &=\int \frac{d^3k}{(2\pi)^{3/2}}W(k/k_{\sigma})\tilde{\rho}_{\mathbf{k}}
	(\tau)e^{-i\mathbf{k}\cdot \mathbf{x}}\,,\;
	\Pi_s(\tau,\mathbf{x})
	=\int \frac{d^3k}{(2\pi)^{3/2}}W(k/k_{\sigma})\tilde{\Pi}_{\mathbf{k}}
	(\tau)e^{-i\mathbf{k}\cdot \mathbf{x}}\,.
	\label{short-splitting}
\end{align}
Note that we have dropped the $\rho$-labels
because in the following we will only be concerned with the
stochastic dynamics of the $\rho$-field.

The Fourier functions in \eqref{short-splitting}
is linear in the mode functions of standard perturbation theory, which are typically kept at linear order in the equations of motion.
(We will not consider non-Gaussianities in this work.)
Following such standard analysis,
we will assume that $\tilde{\rho}_{\mathbf{k}}$
and $\tilde{\Pi_{\mathbf{k}}}$
with $|\mathbf{k}|\gg k_{\sigma}$
solve the linear-order version of
the equation set \eqref{matter-eom-canonical}
when expanded around the background $\bar{\rho}$ and $\bar{\Pi}$.

In particular, this means that if we rewrite the full equations of motion
\eqref{matter-eom-canonical}
for $\rho$-field
into the form (suppressing $\phi$- and $a$-DOFs)
\begin{align}
	\rho'=f(\rho,\Pi)\,,\;\Pi_{\rho}'=g(\rho,\Pi)\,,
	\label{matter-eom-formal}
\end{align}
then $\tilde{\rho}_{\mathbf{k}}$
and $\tilde{\Pi}_{\mathbf{k}}$
with $|\mathbf{k}|\gg k_{\sigma}$
solve
\begin{align}
\tilde{\rho}_{\mathbf{k}}' &=\partial_{\rho}f(\rho,\Pi)|_{\rho=\bar{\rho},\Pi=\bar{\Pi}}
\tilde{\rho}_{\mathbf{k}}
+\partial_{\Pi}f(\rho,\Pi)|_{\rho=\bar{\rho},\Pi=\bar{\Pi}}
	\tilde{\Pi}_{\mathbf{k}}\nonumber\\
\tilde{\Pi}_{\mathbf{k}}' &=\partial_{\rho}g(\rho,\Pi)|_{\rho=\bar{\rho},\Pi=\bar{\Pi}}
\tilde{\rho}_{\mathbf{k}}
+\partial_{\Pi}g(\rho,\Pi)|_{\rho=\bar{\rho},\Pi=\bar{\Pi}}
	\tilde{\Pi}_{\mathbf{k}}\,,
	\label{linear-mode-sol}
\end{align}
where spatial derivatives are understood in Fourier space,
namely $\partial_{\rho}(\partial_i \rho)\tilde{\rho}_{\mathbf{k}}
\rightarrow ik_i \partial_{\rho}(\rho)\tilde{\rho}_{\mathbf{k}}$ and so on.

To obtain the equations of motion for
$\bar{\rho}$ and $\bar{\Pi}$
up to linear order in $\tilde{\rho}_{\mathbf{k}}$
and $\bar{\Pi}_{\mathbf{k}}$ (or equivalently linear in $\rho_s$ and $\Pi_s$), we substitute the splitting \eqref{short-splitting} into
the equations of motion
\eqref{matter-eom-formal}
and use properties \eqref{linear-mode-sol}.
The linear expansion of the LHS and RHS of \eqref{matter-eom-formal}
will partially cancel, leaving behind
\begin{align}
	\bar{\rho}'+\int\frac{d^3k}{(2\pi)^{3/2}}\partial_{\tau}W(k/k_{\sigma})	
	\tilde{\rho}_{\mathbf{k}}e^{-i\mathbf{k}\cdot \mathbf{x}}
	=&
	f(\bar{\rho},\bar{\Pi})\nonumber\\
	\bar{\Pi}'+\int\frac{d^3k}{(2\pi)^{3/2}}\partial_{\tau}W(k/k_{\sigma})
	\tilde{\Pi}_{\mathbf{k}}e^{-i\mathbf{k}\cdot \mathbf{x}}
	=&
	g(\bar{\rho},\bar{\Pi})\,.
	\label{pre-langevin}
\end{align}
A further simplification comes from the fact that
one can ignore any spatial derivatives in
$f(\bar{\rho},\bar{\Pi})$ and $g(\bar{\rho},\bar{\Pi})$.
This is because $\sigma\ll 1$ and
the window function (more precisely $1-W(k/k_\sigma)$) ensures that
$\partial_i\bar{\rho}$ and $\partial_i\bar{\Pi}$
are suppressed by $\sigma$ when the derivatives are applied
to the barred quantities.

Once we apply canonical quantization to $\tilde{\rho}_{\mathbf{k}}$
and $\tilde{\Pi}_{\mathbf{k}}$,
they will follow a quantum distribution (wavefunction),
which can be approximated by classical statistics
in certain cases.
A common realization is to use a Bunch-Davies solution
for $\tilde{\rho}_{\mathbf{k}}$
and $\tilde{\Pi}_{\mathbf{k}}$
so that statistics only rely on two-point functions,
which we can approximate with Gaussian noise terms $\xi_{\rho}$
and $\xi_{\Pi}$ for
\eqref{pre-langevin}:
\begin{align}
	\bar{\rho}' &= f(\bar{\rho},\bar{\Pi})+\xi_{\rho}\nonumber\\
	\bar{\Pi}' &= g(\bar{\rho},\bar{\Pi})+\xi_{\Pi}\nonumber\\
	\xi_{\rho} &=-\int\frac{d^3k}{(2\pi)^{3/2}}\partial_{\tau}W(k/k_{\sigma})
	\tilde{\rho}_{\mathbf{k}}e^{-i\mathbf{k}\cdot \mathbf{x}}\nonumber\\
	\xi_{\Pi} &=-\int\frac{d^3k}{(2\pi)^{3/2}}\partial_{\tau}W(k/k_{\sigma})
	\tilde{\Pi}_{\mathbf{k}}e^{-i\mathbf{k}\cdot \mathbf{x}}~,
	\label{langevin-formal}
\end{align}
where $\tilde{\rho}_{\mathbf{k}}$ and $\tilde{\Pi}_{\mathbf{k}}$
are now understood as stochastic quantities.

Finally,
the canonical version of the stochastic
formalism makes it clear where the
backreaction (or noise) comes from:
Backreaction occurs when
short-wavelength DOFs flow into the
(long-wavelength) background
as they get stretched via exponential
expansion (viz, $\xi\sim \partial_{\tau} W(k/k_{\sigma}$)).

\section{Derivation of equations for the quadratic expectation values}
In the previous appendix, we recalled the stochastic formalism
using the field variable and its conjugate momentum.
In the main text, 
to make the equations of motion simpler notation-wise
(as in \eqref{rho_pi_eqs_L}),
we used instead the rescaled the momentum field
\eqref{rescaled-momentum}.
In this section, we derive the equations of motion
for the quadratic expectation values
of these rescaled variables.

We start by linearizing
the equations of motion of the $\rho$-field in \eqref{rho_EOM_full}.
This is justified if one is interested in the stochastic evolution
around the critical point, where
$\rho$ and $\pi^{\rho}$ are very small. 
Based on the result \eqref{langevin-formal}
of the previous section,
the equations of motion of the $\rho$-field 
(with rescaled momentum $\pi^{\rho}$)
can be written as \eqref{rho_pi_eqs_L}, 
which one can put collectively into the matrix-notation
equation
\begin{equation}\label{Matrix_Gen_Quan}
	{\partial _\tau }{\boldsymbol{\rho }} = {\mathbf{A}}(\tau ){\boldsymbol{\rho }} + {\boldsymbol{\xi }}(\mathbf{x},\tau )~,
\end{equation}
where in our case $\boldsymbol{\rho }=\left(\rho,\pi^{\rho}\right)^{T}$ and ${\boldsymbol{\xi }}(\mathbf{x},\tau )=\left(\xi_{\rho},\xi_{\pi^{\rho}}\right)^T$. If $\tau$ is taken as the efold time, the matrix $\mathbf{A}$ reads
\begin{equation}
	\mathbf{A}=
	\begin{pmatrix}
		&0 &1\\
		&{-H^{-2}m^2_{\tilde\rho,{\rm eff}}\,} &-(3-\epsilon_H)~.
	\end{pmatrix}
\end{equation}

 The solution to \eqref{Matrix_Gen_Quan} with initial condition $\boldsymbol{\rho }_0$ can be derived as \cite{vennin2020stochastic}
 \begin{equation}\label{Gen_Phi_solu}
 	\boldsymbol{\rho }\left( \tau  \right) = {\mathbf{G}}\left( {\tau ,{\tau _0}} \right){\boldsymbol{\rho }_0} + \int_{{\tau _0}}^\tau  d s{\boldsymbol{G}}\left( {\tau ,s} \right){\boldsymbol{\xi }}(s)~,
 \end{equation}
where the Green matrix ${\mathbf{G}}\left( {\tau ,\tau_0} \right)$ can be defined as 
\begin{equation}
	{\mathbf{G}}\left( {\tau ,{\tau _0}} \right)\equiv {\mathbf{U}}\left( \tau  \right){{\mathbf{U}}^{ - 1}}\left( {{\tau _0}} \right)\theta \left( {\tau  - {\tau _0}} \right)~.
\end{equation}
The matrix $\mathbf{U}$ in the definition of Green matrix can be constructed from two independent special solutions of classical EOM $\partial_\tau\boldsymbol{\rho }=\mathbf{A}\boldsymbol{\rho }$ as
\begin{equation}
	{\mathbf{U}}\left( \tau  \right) = 
 (\boldsymbol{\rho }_s^{(1)} ~\boldsymbol{\rho }_s^{(2)})~,
\end{equation}
in which each ${\boldsymbol{\rho }_s}^{(i)}$ represents one special solution of $\partial_\tau\boldsymbol{\rho }=\mathbf{A}\boldsymbol{\rho }$.
We also note that based on the natural requirement
$\boldsymbol{\rho}(\tau_0)=\boldsymbol{\rho}_0$,
we have
$\mathbf{G}(\tau_0,\tau_0)=\mathbf{I}$
and therefore $\theta(0)=1$ for the step-function.
This is in contrast to other common choices
such as $\theta(0)=1/2$ or $0$.
(In our context, the choice matters only for certain integrals
involving delta-functions, which will be demonstrated below.)

For Gaussian noise ${\left\langle 0 \right|} 
 {\boldsymbol{\xi }}\left| 0 \right\rangle  = 0$, the vacuum expectation value of $\boldsymbol{\rho }$ is exactly the one driven by a fully classical process. Therefore, to demonstrate the quantum noise effect, it is necessary to consider the variance of  $\boldsymbol{\rho}$ defined as
 \begin{equation}\label{Gen_Delta}
 	{\mathbf{\Delta }}\equiv \begin{array}{*{20}{c}}
 		{\left\langle 0 \right|} 
 	\end{array}{\boldsymbol{\rho}}{{\boldsymbol{\rho}}^\dag }\left| 0 \right\rangle  - \begin{array}{*{20}{c}}
 		{\left\langle 0 \right|} 
 	\end{array}{\boldsymbol{\rho}}\left| 0 \right\rangle \begin{array}{*{20}{c}}
 		{\left\langle 0 \right|} 
 	\end{array}{{\boldsymbol{\rho}}^\dag }\left| 0 \right\rangle  = \int_{{\tau _0}}^\tau  d s{\mathbf{G}}\left( {\tau ,s} \right){{\mathbf{Z}}}\left( s \right){{\mathbf{G}}^\dag }\left( {\tau ,s} \right).
 \end{equation}
For $\boldsymbol{\rho}=(\rho_1,\rho_2)^T
\equiv (\rho,\pi^{\rho})^T$, Gaussian noise implies $\left\langle 0 \right|{\xi _{{\rho_i}}}\left( {\mathbf x,{\tau _1}} \right){\xi _{{\rho_j}}}\left( {\mathbf x,{\tau _2}} \right)\left| 0 \right\rangle \propto \delta \left(\tau_1-\tau_2\right)$\cite{vennin2020stochastic}. The matrix elements of matrix  $\mathbf{Z }$  in the variance of $\mathbf{\rho}$ are defined through
 \begin{equation}
 		\left\langle 0 \right|{\xi _{{\rho_i}}}\left( {\mathbf x,{\tau _1}} \right){\xi _{{\rho_j}}}\left( {\mathbf x,{\tau _2}} \right)\left| 0 \right\rangle \hfill \equiv {Z _{{\rho_i}{\rho_j}}}\left( {{\tau _1}} \right)\delta \left( {{\tau _1} - {\tau _2}} \right)\,,
 \end{equation}
 which can be found once the mode functions are known.
 
 To derive the evolution of $\mathbf{\Delta}$,
 we recall
 the definition of $\mathbf{U}$ and $\mathbf{G}$. One finds that these matrices satisfy $\displaystyle{\frac{{d{\mathbf{U}}}}{{d\tau }} = {\mathbf{A}}\left( \tau  \right){\mathbf{U}}\left( \tau  \right)}$ and  $\displaystyle{\frac{{\partial {\mathbf{G}}\left( {\tau ,{\tau _0}} \right)}}{{\partial \tau }} = {\mathbf{A}}(\tau ){\mathbf{G}}\left( {\tau ,{\tau _0}} \right) + {\mathbf{I}}\delta \left( {\tau  - {\tau _0}} \right)}$. Therefore, taking the time-derivative of 
 \eqref{Gen_Delta} one obtains
\begin{equation}
	\begin{gathered}
		{\partial _\tau }{\mathbf{\Delta }} = {\partial _\tau }\left( {\int_{{\tau _0}}^\tau  d s{\mathbf{G}}\left( {\tau ,s} \right){\mathbf{Z }}\left( s \right){{\mathbf{G}}^\dag }\left( {\tau ,s} \right)} \right) \hfill \\
		= \int_{{\tau _0}}^\tau  d s\left( {{\partial _\tau }{\mathbf{G}}\left( {\tau ,s} \right){\mathbf{Z }}\left( s \right){{\mathbf{G}}^\dag }\left( {\tau ,s} \right) + {\mathbf{G}}\left( {\tau ,s} \right){\mathbf{Z }}\left( s \right){\partial _\tau }{{\mathbf{G}}^\dag }\left( {\tau ,s} \right)} \right) +  \hfill \\
		\mathop \smallint \nolimits_\tau ^{\tau  + d\tau } ds{\mathbf{G}}\left( {\tau ,s} \right){\mathbf{Z }}\left( s \right){{\mathbf{G}}^\dag }\left( {\tau ,s} \right)/d\tau  \hfill \\
		= \int_{{\tau _0}}^\tau  d s\left( {{\mathbf{A}}\left( \tau  \right){\mathbf{G}}\left( {\tau ,s} \right){\mathbf{Z }}\left( s \right){{\mathbf{G}}^\dag }\left( {\tau ,s} \right) + {\mathbf{G}}\left( {\tau ,s} \right){\mathbf{Z }}\left( s \right){{\mathbf{G}}^\dag }\left( {\tau ,s} \right){{\mathbf{A}}^\dag }\left( \tau  \right)} \right) \hfill \\
  +\int_{\tau_0}^{\tau}\delta(\tau-s)({\mathbf{Z }}\left( s \right){{\mathbf{G}}^\dag(\tau,s) }+\mathbf{G}\left( {\tau ,s} \right){\mathbf{Z }}\left( s \right)) \hfill \\
		+{\mathbf{G}}\left( {\tau ,\tau } \right){\mathbf{Z }}\left( \tau  \right){{\mathbf{G}}^\dag }\left( {\tau ,\tau } \right)  \hfill \\
		= {\mathbf{A}}\left( \tau  \right){\mathbf{\Delta }}\left( \tau  \right) + {\mathbf{\Delta }}\left( \tau  \right){{\mathbf{A}}^\dag }\left( \tau  \right) + {\mathbf{Z }}\left( \tau  \right)~. \hfill \\ 
	\end{gathered} 
 \label{Delta-EOM}
\end{equation}
To get the last equality, we used the identity
associated with choosing $\theta(0)=1$
\begin{align*}
    \int^{\tau}_{\tau_0}dsf(s)\delta(\tau-s)
    =&\int^{\tau}_{\tau_0}dsf(s)(-\partial_s\theta(\tau-s))\\
    =&\int^{\tau}_{\tau_0}ds \partial_sf(s)
    \theta(\tau-s)-[f(\tau)\theta(\tau-\tau)-f(\tau_0)\theta(\tau-\tau_0)]\\
    =&f(\tau)-f(\tau_0)-[f(\tau)-f(\tau_0)]=0\,.
\end{align*}

Using definition \eqref{Gen_Delta},
the evolution equation
\eqref{Delta-EOM} can be equivalently expressed as
\begin{equation}\label{Gen_EoM_delta}
	{\partial _\tau }\left\langle {{\mathbf{\Phi }}{{\mathbf{\Phi }}^\dag }} \right\rangle  = {\mathbf{A}}(\tau )\left\langle {{\mathbf{\Phi }}{{\mathbf{\Phi }}^\dag }} \right\rangle  + \left\langle {{\mathbf{\Phi }}{{\mathbf{\Phi }}^\dag }} \right\rangle {{\mathbf{A}}^\dag }(\tau ) + {\mathbf{Z }}\left( \tau  \right)~.
\end{equation}
By using our two-field equations of motion \eqref{rho_pi_eqs_L} along with \eqref{Gen_EoM_delta}, we obtain equations of motion \eqref{FP_eqs} for the quadratic expectation values.

\clearpage

\providecommand{\href}[2]{#2}\begingroup\raggedright\endgroup

\end{document}